\documentclass[10pt]{IEEEtran}
\usepackage{amsmath}
\usepackage{amssymb}
\usepackage{amsfonts}
\newcommand{\Rmnum}[1]{\expandafter\@slowromancap\romannumeral #1@}
\usepackage{graphicx}
\usepackage{epsfig}
\usepackage{subfigure}
\usepackage[sort]{cite}
\usepackage{xcolor}
\usepackage{enumerate}
\usepackage{extarrows}
\usepackage{algorithmic}
\usepackage{subfigure}
\usepackage[lined,ruled,linesnumbered]{algorithm2e}
\usepackage{tensor}
\usepackage[colorlinks,linkcolor=blue,anchorcolor=blue,citecolor=blue,urlcolor=black]{hyperref}
\newcommand{\rev}[1]{{\color{black}#1}}

\begin{document}
\title{Non-Coherent Over-the-Air Federated Learning: Protocol, Convergence, and Device Scheduling}

\author{Haifeng Wen, Nicolò Michelusi, Osvaldo Simeone, Yang Yang, and Hong Xing
\thanks{H. Wen and H. Xing are with the IoT Thrust, The Hong Kong University of Science and Technology (Guangzhou), Guangzhou, 511453, China; 
H. Xing is also affiliated with the Department of ECE, The Hong Kong University of Science and Technology, HK SAR (e-mails: hwen904@connect.hkust-gz.edu.cn,~hongxing@ust.hk). 
Nicolò Michelusi is with the School of Electrical, Computer and Energy Engineering, Arizona State University, Tempe, AZ 85281 USA (e-mail: nicolo.michelusi@asu.edu).
O. Simeone is with the Institute for Intelligent Networked Systems, Northeastern University London, London, E1 8PH, UK (email: o.simeone@nulondon.ac.uk).
Yang Yang is with the Shanghai HKU Education Center, Shanghai 201210, China (email: yang2@hku.hk) 

A preliminary version of this work appeared in IEEE International Conference on Communications (ICC), QC, Canada, Jun. 2025 \cite{wen2025icc}.
}
}
\maketitle

\begin{abstract}
To mitigate the scalability bottleneck in the radio access network (RAN) in federated edge learning (FEEL), over-the-air federated learning (AirFL) exploits waveform superposition over multiple-access channels (MACs) for analog model aggregation. 
However, coherent AirFL typically relies on stringent PHY-layer conditions such as accurate channel state information (CSI), tight time/frequency synchronization, and frequent transceiver calibration for signal alignment. 
However, these requirements, if not impossible to be met, incur substantial communication and computation overhead.
In this paper, we propose a {\it non-coherent AirFL (NCAirFL)} protocol over a broadband single-antenna MAC, leveraging binary dithering, unbiased non-coherent detection, and long-term error feedback to waive the need for instantaneous CSI. 
For NCAirFL with general smooth non-convex objectives and a constant learning rate, we establish a convergence bound achieving the convergence rate in the same order of $\mathcal{O}(1/\sqrt{T})$ as communication-ideal FedAvg, where $T$ is the total number of communication rounds.
To further improve communication efficiency under data and wireless resource heterogeneity, we also derive a lower bound on the expected single-round objective decrease in the global loss conditioned on device scheduling, building upon which a surrogate objective function is obtained for jointly optimal device selection and power control.
Experimental results on MNIST and CIFAR-10 corroborate that NCAirFL achieves learning performance close to FedAvg in practical settings, with the proposed device scheduling policy substantially accelerating convergence.
\end{abstract}


\IEEEpeerreviewmaketitle
\newtheorem{definition}{\underline{Definition}}[section]
\newtheorem{fact}{Fact}
\newtheorem{assumption}{Assumption}
\newtheorem{theorem}{\underline{Theorem}}[section]
\newtheorem{lemma}{\underline{Lemma}}[section]
\newtheorem{proposition}{\underline{Proposition}}[section]
\newtheorem{corollary}[proposition]{\underline{Corollary}}
\newtheorem{example}{\underline{Example}}[section]
\newtheorem{remark}{\underline{Remark}}[section]
\newcommand{\mv}[1]{\boldsymbol{#1}}
\newcommand{\mb}[1]{\mathbb{#1}}
\newcommand{\Myfrac}[2]{\ensuremath{#1\mathord{\left/\right.\kern-\nulldelimiterspace}#2}}
\newcommand\Perms[2]{\tensor[^{#2}]P{_{#1}}}

\section{Introduction}

The evolution towards artificial intelligence (AI)-native radio access network (RAN) is expected to make learning, inference, and control integrated components of the 6G network infrastructure \cite{ITU-R2023Framework}. In this context, federated learning (FL), popularized by FedAvg \cite{mcmahan2017communication}, has emerged as a privacy-preserving distributed learning framework, whereby devices perform local training and only exchange model updates instead of raw data with a coordinating server. 
This architecture is of particular interest for data-driven and privacy-aware edge applications like healthcare and autonomous systems  \cite{Tao2024Federated}, which paves the way for {\it federated edge learning (FEEL)}, thus enabling model training over volumes of data across Internet-of-Things (IoT) devices such as sensors, mobile terminals, and autonomous vehicles.

However, FEEL inherits several fundamental challenges, most notably, e.g., communication bottlenecks, statistical heterogeneity across local datasets, and resource heterogeneity across devices \cite{Tao2024Federated}.
Among these challenges, the communication bottleneck is especially critical because modern models require repeated exchange of high-dimension updates over limited wireless channels \cite{mcmahan2017communication}. Over-the-air federated learning (AirFL) addresses this issue by exploiting the superposition property of the wireless multiple-access channel (MAC) so that multiple devices can transmit simultaneously, followed by direct recovery of the aggregated update in the analog domain at the server \cite{nazer2007computation,zhu2019broadband}. 
Despite drastically reducing bandwidth from \(\mathcal O(n)\) to \(\mathcal O(1)\), where $n$ denotes the number of participating devices, AirFL couples wireless impairments and learning dynamics much more tightly than conventional FEEL \cite{zhu2019broadband,Tao2024Federated}. 
% In particular, aggregation distortion depends on channel fading, noise, device participation, and power control, and these factors directly affect the optimization trajectory and the final learning performance \cite{zhu2019broadband,Tao2024Federated}.
As a result, most existing AirFL protocols adopt coherent detection by, e.g., equalization, beamforming, and scheduling, for analog aggregation to ensure alignment of amplitudes and/or phases across distributed devices \cite{zhu2019broadband,yang2020federated,amiri2020federated}.

Coherent-detection-based aggregation nevertheless requires not only accurate channel state information (CSI) acquisition but also tight time/frequency synchronization across devices and stringent calibrations over radio-frequency (RF) chains, which are, if not impossible, particularly demanding for broadband MAC, where CSI estimation, synchronization, and calibration incur a considerable amount of overhead \cite{tegin2023federated,Tao2024Federated}.
Besides, channel-inversion-based equalization is intrinsically sensitive to deep fades, which amplifies effective noise or forces aggressive device truncation, thereby causing a non-trivial trade-off between aggregation quality and data exploitation \cite{zhu2019broadband,wen2024AirFL-Mem}. 
In addition, data and wireless resource heterogeneity worsen such influence in training trajectory, calling for joint device selection and power control \cite{su2022data,sun2024channel}
% \cite{su2022data,sun2024channel,nguyen2021fast,zhang2022communication,amiri2021convergence}.
Motivated by the above limitation of coherent-based aggregation, we propose in this paper a {\it non-coherent} detection-based broadband AirFL training scheme, namely, {\it NCAirFL}. NCAirFL approximates the learning performance of communication-ideal FedAvg with theoretical guarantee, and enjoys practical benefits of waiving instantaneous CSI acquisition (just leaving large-scale CSI estimation), i.e., {\it CSI semi-free}.

\subsection{Related Works} 
Building on the over-the-air computing principle in \cite{nazer2007computation}, a substantial line of work has investigated AirFL to reduce uplink latency by replacing orthogonal transmissions over MAC with analog aggregation. 
The representative broadband implementation in \cite{zhu2019broadband} used truncated channel inversion (TCI) for analog aggregation over orthogonal frequency division multiplexing (OFDM), and reveals explicit trade-offs among receive SNR, aggregation reliability, and the fraction of participating devices. 
The MIMO extension in \cite{yang2020federated} jointly designed device selection and beamforming to reduce aggregation error. The fading MAC schemes in \cite{amiri2020federated} included a compressed analog DSGD design based on gradient sparsification, dimensionality reduction, error accumulation, and power control.
More recently, the coherent TCI-based scheme in \cite{wen2024AirFL-Mem} showed that a long-term memory mechanism can recover the communication-learning trade-off lost in deep fading scenarios, and match the convergence-rate order of ideal FedAvg.
\cite{sery2020analog} proposed a gradient-based multiple access protocol for AirFL, in which devices compensate for the channel phase only.
% \cite{xia2021fast} proposed an over-the-air-based FedSplit method, which combines operator-splitting-based local proximal updates with channel inversion to achieve a fast convergence rate.
These works collectively demonstrate the promise of AirFL, but they also make it clear that coherent detection-based analog aggregation remains strongly dependent on CSI acquisition.

A parallel line of research has therefore sought to reduce or remove CSI requirements. In the broader AirComp literature, CSI-efficient or blind aggregation methods have been investigated from a signal-processing perspective, including receive-side design for over-the-air computation and blind recovery methods for unknown channels \cite{deng2025robust,chen2018over,dong2020blind}. 
However, the performance metric in AirFL is ultimately the end-to-end learning behavior, including convergence speed and final model quality \cite{cao2021optimized,Tao2024Federated}. 
This has motivated learning-oriented CSI-light designs. 
Blind federated edge learning in \cite{amiri2021blind} assumed no CSI at transmitters (TXs) and imperfect CSI at a multi-antenna receiver (Rx), and used convergence analysis to quantify the role of the number of antennas.
The work \cite{yang2021revisiting} studied analog AirFL in the presence of interference and shows that interference can simultaneously hinder convergence and improve generalization, thereby highlighting that communication impairments may affect learning in more nuanced ways~\cite{wen2026tccn}. 
The random-orthogonalization framework in \cite{wei2023random} and the random-access framework in \cite{choi2022communication} exploited channel hardening and favorable propagation in massive MIMO systems to enable natural over-the-air aggregation without uplink transmitter-side CSI, thus substantially reducing estimation overhead at the receiver. 
% In another random-access formulation, \cite{choi2022communication} developed a scalable distributed-SGD scheme whose iteration time is essentially independent of the number of devices and that can employ non-coherent combining over fading channels. 
However, massive MIMO-based systems introduce significant complexity and deployment costs.

Closest in spirit to our work, \cite{michelusi2024non} developed a non-coherent over-the-air gradient method for fully \emph{decentralized} learning that does not require MIMO, scheduling, topology information, or CSI, and provides convergence guarantees with \emph{strongly convex} objectives leveraging energy superposition and unbiased non-coherent consensus estimation.
The extension in \cite{11587673} further developed this non-coherent principle to improve robustness against interference.
% The related method in \cite{deng2025robust} uses non-coherent energy detection for over-the-air data aggregation, and develops a decentralized projected-gradient framework with joint transmit--receive power scaling to eliminate stochastic consensus bias. 
These two works addressed over-the-air decentralized optimization or consensus under communication models and objective classes that differ from the broadband AirFL with general \emph{smooth and non-convex} objectives considered herein. 
% Our focus is to establish a single-antenna non-coherent AirFL construction and convergence guarantees for non-convex objectives under the specific architecture studied in this paper.
% A CSI semi-free \emph{broadband, single-antenna} AirFL design with rigorous non-convex convergence guarantees, therefore, remains largely open.
% \nm{this last sentence seems a bit repetitive with the previous one..}

Another key ingredient in AirFL is device selection and resource allocation. 
As a result of data and wireless resource heterogeneity, selecting devices solely according to channel quality may be far from optimal.
The framework in \cite{11475389} allows controlled aggregation bias to mitigate the bottleneck caused by poor device channels, and jointly optimizes the resulting bias--variance trade-off for over-the-air and digital FL.
The importance and channel-aware rule in \cite{ren2020scheduling} uses gradient divergence and probabilistic aggregation to preserve unbiasedness. The design in \cite{yue2022efficient} adopts a related one-round learning-performance perspective for federated meta-learning over wireless multiple-access links. 
% Outside the strict AirFL setting, the FOLB scheme in \cite{nguyen2021fast} uses lower bounds on the round-wise loss decrease to guide intelligent client sampling, thereby accelerating convergence in heterogeneous FL systems. Probabilistic device participation is optimized in \cite{zhang2022communication} to reduce overall communication time while preserving unbiased aggregation.
% For wireless FL with orthogonal or scheduled access, update-aware scheduling and resource allocation were studied in \cite{amiri2021convergence}, where transmission decisions depend jointly on channel states and the significance of local model updates. 
As for AirFL specifically, \cite{su2022data} developed a dynamic data and channel-adaptive scheduling along with a power-control strategy leveraging residual feedback and Lyapunov-drift optimization, while \cite{sun2024channel} proposed a probabilistic over-the-air scheduling framework whose convergence analysis explicitly links device selection to both communication distortion and update variance. Related gradient and channel-aware dynamic scheduling ideas were further explored in \cite{du2023gradient}. 
These results together underline the value of learning-aware participation control, but they are not aimed at the non-coherent AirFL studied in this paper.

\begin{figure*}[t]
    \centering
    \includegraphics[width=1\linewidth]{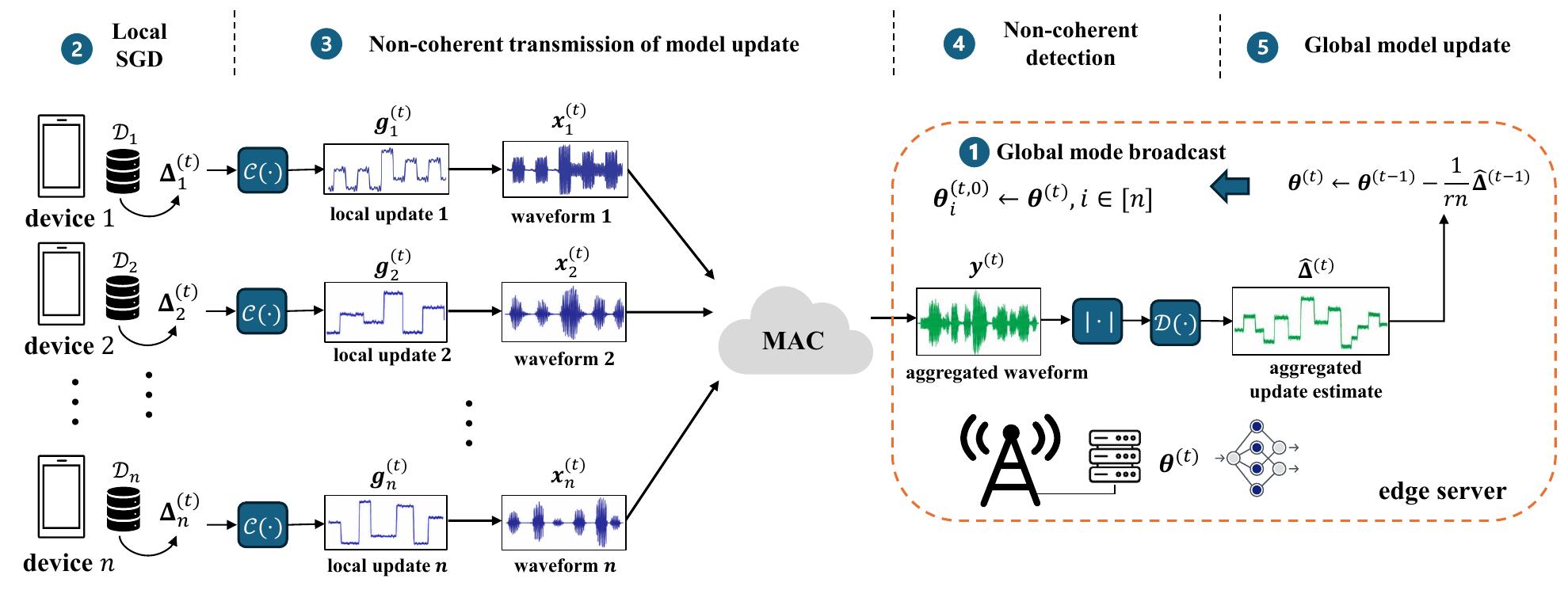}
    \caption{Illustration of the considered non-coherent AirFL system. Here, $\mathcal{C}(\cdot)$ and $\mathcal{D}(\cdot)$ denote the preprocessing and decoding mappings for non-coherent transmission, respectively, and $|\cdot|^2$ denotes square-law detection. In each communication round, all devices receive the global model from the broadcast and overwrite their local models. 
    Under optimized selection, all devices perform local SGD updates, compute candidate local model innovations, and report two scalar statistics, while only the selected devices transmit their preprocessed innovations via $\mathcal{C}(\cdot)$ over the multiple-access channel (MAC). 
    The edge server receives the superimposed signal, applies square-law detection and the decoder $\mathcal{D}(\cdot)$ to estimate the average model innovations and updates the global model.}
    \label{fig:system model}
\end{figure*}

\subsection{Contributions}
This paper investigates a communication-efficient CSI semi-free broadband AirFL architecture with the goal of achieving theoretically guaranteed learning performance as communication-ideal FedAvg and practical convergence acceleration. The main contributions are summarized as follows.
\begin{enumerate}
    \item Building on the non-coherent single-antenna receiving design developed in \cite{michelusi2024non}, we propose a CSI semi-free broadband AirFL protocol, \emph{NCAirFL}, which introduces binary dithering, unbiased non-coherent detection, and a long-term error-feedback mechanism to approximate communication-ideal FedAvg.
    \item For general $L$-smooth non-convex objectives, we analyze the convergence rate of NCAirFL, which, with a constant learning rate, scales at the same order as FedAvg.
    \item Inspired by the established convergence, we derive a lower bound on the expected single-round decrease in global loss conditioned on device scheduling, and use this bound as a surrogate objective for joint device selection and power control, which is solved optimally with low cost.
    \item Experiments verify that NCAirFL can approach the performance of communication-ideal FedAvg in practical settings, and that the jointly optimal device-selection and power-control policy can further accelerate convergence compared to other uniform-selection baselines.
\end{enumerate}

\section{System Model}\label{sec:System Model}
In this section, we first review the basic FL protocol, FedAvg, which operates under the assumption of ideal and noiseless communications. Then, we present the considered broadband channel model~\cite{zhu2019broadband}, as illustrated in Fig.~\ref{fig:system model}. 
A set $[n]\triangleq\{1,\ldots,n\}$ of devices communicates with an edge server over a broadband multiple access fading channel via analog OFDM, in which neither the transmitting devices nor the edge server has access to instantaneous CSI.

\subsection{Learning Protocol (FedAvg)}\label{subsec:Learning Model}

As illustrated in Fig.~\ref{fig:system model}, each device $i\in[n]$ holds a local dataset $\mathcal D_i=\{\mv\xi_{i,j}\}_{j\in[m]}$ containing $m$ data samples.
The $n$ devices collaboratively train a shared machine learning model parameterized by $\mv \theta\in\mathbb{R}^d$ to solve an empirical loss minimization problem given by 
\begin{align*}
\mathrm{(P0)}:\qquad \operatorname*{\mathtt{Minimize}}_{\mv \theta \in \mathbb{R}^d} \quad f(\mv \theta)\triangleq\frac{1}{n}\sum_{i=1}^{n}f_{i}(\mv \theta).
\end{align*}
In $\rm{(P0)}$, $f:\mathbb{R}^d\to[0,+\infty)$ is the global empirical loss function, while \(f_{i}(\mv \theta)=\Myfrac{1}{m}\sum_{\boldsymbol\xi\in\mathcal{D}_{i}}\ell(\mv \theta;\boldsymbol\xi)\) represents the local empirical loss function at device $i\in[n]$. 
Because $|\mathcal D_i|=m$ for every device, uniform device averaging in $\rm{(P0)}$ is equivalent to uniform averaging over all samples. The sample-wise loss $\ell(\mv \theta;\boldsymbol\xi)$ is evaluated at model $\mv\theta$ and data sample $\mv\xi$.
% Since the datasets $\{\mathcal{D}_i\}_{i=1}^n$ are privately distributed across devices, the devices must collaborate to solve \(\mathrm{(P0)}\) without sharing their raw data.

To solve \(\mathrm{(P0)}\), we adopt the standard FedAvg protocol in~\cite{mcmahan2017communication}. 
In each communication round, each (selected) device in $[n]$ initializes its local model using the current global model, performs multiple local SGD steps using its own dataset, and then transmits the local model innovation to the server. The server aggregates the received model innovations to update the global model, which is subsequently broadcast to all devices for the next round. 
% \rev{Under the optimized policy in Section~\ref{sec:Optimization}, all devices compute candidate local updates before selection, but only the selected devices transmit their innovations.}
This procedure continues until a prescribed number of communication rounds, denoted by \(T\), is reached, or a prescribed stopping criterion is satisfied.

Specifically, at communication round \(t\in\{0,1,\ldots,T-1\}\), \rev{a general device-selection policy (optimized in Sec. \ref{sec:Optimization}) chooses an active set \(\mathcal{I}^{(t)}\subseteq[n]\). Let $s\in[n]$ denote the number of active devices, so that $|\mathcal I^{(t)}|=s=rn$, where $r$ is the device participation ratio.}  
\rev{Each device \(i\in [n]\)} initializes its local model as \(\mv \theta_i^{(t,0)}\leftarrow \mv \theta^{(t)}\), and then performs \(Q\) local SGD steps using mini-batches \(\mathcal{B}_i^{(t,q)}\subseteq \mathcal{D}_i\)\footnote{Under uniform selection, only active devices perform local update. Under the optimized policy in Section~\ref{sec:Optimization}, all devices compute candidate local updates and upload two scalar statistics before the active device set is selected.}:
\begin{equation} \label{eq:local updates}
    \boldsymbol \theta_{i}^{(t, q+1)} \leftarrow \boldsymbol \theta_{i}^{(t, q)}-\rev{\eta}\hat\nabla f_{i}(\boldsymbol \theta_{i}^{(t, q)}),
\end{equation}
where \(q\in\{0,\ldots,Q-1\}\) is the local iteration index; \rev{$\eta>0$ is a constant learning rate}; and \(\hat\nabla f_{i}(\boldsymbol \theta_{i}^{(t,q)})\) is the stochastic gradient vector evaluated on the mini-batch as
\begin{equation}
    \hat\nabla f_{i}(\boldsymbol \theta_{i}^{(t, q)})=
    \frac{1}{|\mathcal{B}_{i}^{(t,q)}|}\sum_{\boldsymbol\xi\in\mathcal{B}_{i}^{(t,q)}}\nabla \ell (\boldsymbol \theta_{i}^{(t, q)};\boldsymbol\xi),
\end{equation}
where the mini-batches are sampled uniformly and independently across devices, local iterations, and communication rounds.

After completing the local updates, device \(i\in [n]\) obtains the model innovation
\(
\mv{\Delta}_{i}^{(t)} = \mv \theta_{i}^{(t,Q)}-\mv \theta_{i}^{(t,0)}.
\)

Under ideal communication, the server would update the global model by averaging the model innovations over all active devices:
\begin{equation}\label{eq:global updates}
    \mv \theta^{(t+1)} \leftarrow \mv \theta^{(t)} + \frac{1}{s}\sum_{i \in \mathcal{I}^{(t)}}\mv{\Delta}_{i}^{(t)}.
\end{equation}
The ideal average in \eqref{eq:global updates} is not directly available over a wireless channel due to fading and noise. This motivates the communication model and non-coherent aggregation described next.

\subsection{Communication Model} \label{subsec:Communication Model}
At round \(t\), the devices in \(\mathcal{I}^{(t)}\) simultaneously transmit vectors \(\mv x_i^{(t)}=[x_{i,1}^{(t)},\ldots,x_{i,d}^{(t)}]^T\) over the same $d$ OFDM subcarriers. The received signal at the server is~\cite{zhu2019broadband}
\begin{equation} \label{eq:received signal}
    \boldsymbol y^{(t)}=\sum_{i\in\mathcal{I}^{(t)}} \sqrt{\kappa_i}\boldsymbol h_{i}^{(t)}\odot \boldsymbol x_{i}^{(t)} + \boldsymbol n^{(t)},
\end{equation}
where $\kappa_i$ is the large-scale fading power gain, \(\boldsymbol h_i^{(t)}=[h_{i,1}^{(t)},\ldots,h_{i,d}^{(t)}]^T\) is the small-scale fading vector for device \(i\), \(\boldsymbol n^{(t)}\sim\mathcal{CN}(\mv 0,\sigma^2\mv I_d)\) is additive Gaussian noise, and \(\odot\) denotes the Hadamard product.
For every $i$ and $j$, $h_{i,j}^{(t)}$ is a proper complex random variable with zero mean, unit variance, i.e., $\mathbb E[|h_{i,j}^{(t)}|^2]=1$, and finite fourth moment satisfying
\(
\mathbb{E}[(| h_{i,j}^{(t)}|^2 -1 )^2] = M_h < \infty.
\)
If \(h_{i,j}^{(t)}\) is circularly symmetric complex Gaussian, then \(M_h=1\). The vector \(\mv h_i^{(t)}\) remains constant within round \(t\), but may vary across rounds.\footnote{If the physical channel is static, one may instead introduce random phase shifts and coordinated circular subcarrier shifts at the transmitters to satisfy the above assumption~\cite{michelusi2024non}.}
The fading vectors are assumed independent across devices and communication rounds. No independence is required across OFDM subcarriers of the same device.
% \nm{i think independence assumption should be stated here, after the block fading model. IT feels more logical.}

We next specify the transmitted vectors. Since instantaneous CSI is unavailable, the considered non-coherent architecture uses neither small-scale channel inversion nor coherent phase alignment. Let \(\mathcal{C}:\mathbb{R}^d\to\mathbb{R}_+^d\) map the local model innovation \(\mv \Delta_i^{(t)}\) to a nonnegative vector \(\mv g_i^{(t)}=\mathcal{C}(\mv \Delta_i^{(t)})\). The mapping is specified in the next section. For each active device, the transmitted vector is constructed entry-wise as
\begin{equation} \label{eq:non-coherent transmitted signal}
    x_{i,j}^{(t)} = \alpha_i^{(t)}\sqrt{\Myfrac{g_{i,j}^{(t)}}{\rev{\eta}}},\qquad j\in[d],
\end{equation}
where \(g_{i,j}^{(t)}\) is the \(j\)-th entry of \(\mv g_i^{(t)}\), \(j \in [d]\); and \(\alpha_i^{(t)}\) is a power scaling factor chosen to only equalize the large-scale fading coefficient. 
We choose
\[
\alpha_i^{(t)}=\Myfrac{\sqrt{\rho^{(t)}}}{\sqrt{\kappa_i}},
\]
where $\rho^{(t)}>0$ is the common power scaling coefficient in round $t$ and is selected so that each active device satisfies the following per-round average transmit-power constraint: \cite{sery2021over,amiri2020federated,cao2021optimized}
\begin{equation} \label{eq:power constraint}
    \frac{1}{d} \mathbb{E}\!\left[\|\boldsymbol x_{i}^{(t)}\|^2\,\middle|\,\rev{\mathcal F_0}\right] \le P_i,
    \qquad i\in\mathcal I^{(t)}.
\end{equation}
Here, $\mathcal F_0$ denotes the initialization $\sigma$-algebra defined in Section~\ref{sec:Convergence Analysis}, The sequences $\{\mathcal I^{(t)},\rho^{(t)}\}_{t=0}^{T-1}$ used in the analysis may be initialized either randomly or deterministically, but they are $\mathcal F_0$-measurable and are not adapted to the learning trajectory. Thus, conditional on which the expectation in \eqref{eq:power constraint} averages over only the learning and communication randomness. 
% The largest feasible $\rho^{(t)}$ may consequently be random through its initialization-time dependence on $\mathcal I^{(t)}$.

% \rev{For vanilla NCAirFL and the convergence analysis, we use the largest common scaling that satisfies \eqref{eq:power constraint}, namely}
% \begin{equation}\label{eq:maximum feasible rho}
% \rev{\rho^{(t)}=\min_{i\in\mathcal I^{(t)}}
% \frac{d\kappa_i\eta^{(t)}P_i}{\mathbb E\left[\|\mv g_i^{(t)}\|_1\right]}.}
% \end{equation}
% \hf{We intentionally use an unconditional per-round average-power constraint in \eqref{eq:power constraint}, rather than an instantaneous power constraint, as in \cite{sery2021over,amiri2020federated,cao2021optimized}. The expectation averages over the learning history and current-round randomness. This choice is consistent with the unconditional moment bounds used in the convergence and one-round analyses. Otherwise, we need almost-sure bounds for the norms of stochastic gradient and memory, which are too strong.}

Substituting the above choice of \(\alpha_i^{(t)}\) into \eqref{eq:received signal}, the \(j\)-th entry of the received signal is
\begin{equation} \label{eq:simplified received signal}
    y_j^{(t)}= \sqrt{\rho^{(t)}} \sum_{i\in\mathcal{I}^{(t)}} h_{i,j}^{(t)} \sqrt{ \Myfrac{g_{i,j}^{(t)}}{\rev{\eta}}} + n_j^{(t)}.
\end{equation}
Hence, unlike coherent AirFL, the present design compensates only for large-scale fading and does not require compensation for small-scale fading~\cite{zhu2019broadband,amiri2020federated}.

After receiving \eqref{eq:simplified received signal}, the server aims to recover the superimposed signal $\sum_{i\in\mathcal I^{(t)}}\mv g_i^{(t)}$ without access to CSI.
For this purpose, we employ square-law detection on each subcarrier \(j \in [d]\) and define the statistic
\begin{equation} \label{eq:NC receiver}
    r_j^{(t)} = \Myfrac{(|y_j^{(t)}|^2-\sigma^2)}{\rho^{(t)}}.
\end{equation}
Conditional on the transmitted quantities, \(r_j^{(t)}\) is an unbiased estimate of \(\sum_{i\in \mathcal{I}^{(t)}} g^{(t)}_{i,j}/\rev{\eta}\).

Finally, the server applies a decoder \(\mathcal{D}:\mathbb{R}^d\to\mathbb{R}^d\) to \(\mv r^{(t)}\) and obtains an estimate of the aggregate model innovation:
\[
\widehat{\mv \Delta}^{(t)}=\mathcal{D}(\mv r^{(t)}).
\]
The specific form of \(\mathcal{D}\) will be introduced in the next section. Using \(\widehat{\mv \Delta}^{(t)}\), the server updates the global model according to
\begin{equation}
    \mv \theta^{(t+1)} \leftarrow \mv \theta^{(t)}+\frac{1}{s}\widehat{\mv \Delta}^{(t)}.
    \label{eq:airfl global update}
\end{equation}
Therefore, the remaining challenge is to design $\mathcal C(\cdot)$ and $\mathcal D(\cdot)$ so that the detector is unbiased for the preprocessed sum and has controlled estimation error.

\section{Non-coherent AirFL (NCAirFL)}

In this section, we develop a CSI semi-free and communication-efficient AirFL scheme, referred to as \emph{non-coherent AirFL (NCAirFL)}. The main challenge is to design the preprocessing and decoding functions, denoted by $\mathcal{C}(\cdot)$ and $\mathcal{D}(\cdot)$, for the non-coherent transmission model in Section~\ref{subsec:Communication Model}, such that wireless aggregation can be carried out without CSI while preserving learning performance close to that of ideal FedAvg.

The proposed NCAirFL scheme is based on \emph{multiplicative binary dithering}. 
Each device maintains a memory vector $\mv m_i^{(t)}$, initialized as $\mv 0$, that stores the accumulated portion of its past model innovations not represented in previous transmissions. For each candidate device $i\in[n]$, the preprocessing function $\mv g_i^{(t)}=\mathcal{C}(\mv \Delta_i^{(t)})$ is defined entry-wise as
\begin{equation} \label{eq:sparsified model different}
    g_{i,j}^{(t)} = \max \left(\left(m_{i,j}^{(t)}+\Delta_{i,j}^{(t)}\right) \phi_j^{(t)}, 0 \right),
\end{equation}
where  $m_{i,j}^{(t)}$ and $\Delta_{i,j}^{(t)}$ are the $j$th entries of a memory vector $\mv m_i^{(t)}$ and $\mv \Delta_i^{(t)}$, respectively; $j \in [d]$; and $\mv \phi^{(t)}=[\phi_1^{(t)},\ldots,\phi_d^{(t)}]^T$ is a binary random vector whose entries are i.i.d. according to
\begin{equation} \label{eq:flipping vector}
\Pr(\phi = 1) = p, \quad \Pr(\phi = -1) = 1-p
\end{equation}
with $0<p<1$.
The operation in \eqref{eq:sparsified model different} maps the dithered model innovation plus memory into a nonnegative vector suitable for non-coherent transmission.

For a selected device, the memory is updated by subtracting the conveyed surrogate $\mv\phi^{(t)}\odot\mv g_i^{(t)}$ from the sum of the current innovation and previous residual, while an unselected device retains its memory, i.e.,
\begin{equation} \label{eq:memory update rule}
\boldsymbol m_i^{(t+1)}=
\begin{cases}
    \boldsymbol m_i^{(t)}+\boldsymbol \Delta_i^{(t)} - \boldsymbol \phi^{(t)}\odot\boldsymbol g_i^{(t)} & \text{if } i\in\mathcal{I}^{(t)}, \\ 
    \boldsymbol m_i^{(t)} & \text{otherwise}.
\end{cases}
\end{equation}
The memory therefore acts as error feedback. It carries forward the portion of the innovation that the one-sided nonnegative preprocessing does not convey in the current round. Near a stationary point, local innovations become smaller, and the contraction property established in Lemma~\ref{lemma:contraction} then ensures that the accumulated residual diminishes in mean square.

The binary dither vector $\mv \phi^{(t)}$ is generated independently in each round using a pseudorandom generator with its seed shared \emph{a priori} by all devices and the server. 
As a result, the server can remove the effect of the dither during decoding. Specifically, the decoding function $\widehat{\mv \Delta}^{(t)}=\mathcal{D}(\mv r^{(t)})$ is defined as
\begin{equation} \label{eq:approximate aggregated model different}
    \widehat{\mv \Delta}^{(t)} = \rev{\eta} \mv\phi^{(t)}\odot\mv r^{(t)}.
\end{equation}
Substituting \eqref{eq:approximate aggregated model different} into \eqref{eq:airfl global update} yields the global update rule
\begin{equation} \label{eq:NCAirFL global update}
    \boldsymbol \theta^{(t+1)} \leftarrow \boldsymbol \theta^{(t)} + \frac{\rev{\eta}}{s}\boldsymbol \phi^{(t)}\odot \boldsymbol r^{(t)}.
\end{equation}
Algorithm~\ref{alg:Algorithm 1} summarizes the full procedure. First, all devices initialize from the broadcast model and compute model innovations \rev{(lines 6-10)}. When optimized selection is enabled, they also preprocess the innovations \rev{(lines 11)} and report their two scalar norm statistics \rev{(line 12)}. The server then determines $\mathcal I^{(t)}$ and $\rho^{(t)}$ \rev{(line 14)}. Only active devices commit the memory update and transmit (lines 15-18). Finally, the server applies square-law detection, removes the shared dither, updates the global model, and broadcasts it \rev{(c.f.~lines 19-24)}.
% \nm{I would refer to specific line of the algorithm as well.}

\begin{algorithm}[t] 
\SetKwInOut{Input}{Input}
\SetKwInOut{Output}{Output}
\SetKwBlock{DeviceParallel}{On all devices $i \in [n]$ (in parallel):}{end}
\SetKwBlock{ActiveParallel}{On active devices $i \in \mathcal{I}^{(t)}$ (in parallel):}{end}
\SetKwBlock{localSGD}{for $q=0$ to $Q-1$ do}{end}
\SetKwBlock{OnServer}{On server:}{end}
\caption{NCAirFL} \label{alg:Algorithm 1}
\textbf{Input:} learning rate $\eta$, power constraints $\{P_i\}_{i\in[n]}$, number of communication rounds $T$, number of local SGD steps $Q$, active device ratio $r$, and a shared pseudorandom seed for generating $\{\mv \phi^{(t)}\}$ \\
Initialize $\mv m_i^{(0)}=\mv 0$ for all $i\in[n]$, set $t=0$, and initialize the global model $\mv \theta^{(0)}$ \\
\While{$t < T$}{
Generate $\mv \phi^{(t)}$ according to \eqref{eq:flipping vector}\;
\DeviceParallel{
$\mv \theta_i^{(t,0)} \leftarrow \mv \theta^{(t)}$\;
\localSGD{
$\mv \theta_i^{(t,q+1)} \leftarrow \mv \theta_i^{(t,q)} - \rev{\eta} \hat\nabla f_i(\mv \theta_i^{(t,q)})$\;
}
$\mv \Delta_i^{(t)} \leftarrow \mv \theta_i^{(t,Q)}-\mv \theta_i^{(t,0)}$\;
Compute $g_{i,j}^{(t)}$ via \eqref{eq:sparsified model different} for all $j\in[d]$\;
If Algorithm~\ref{alg:Algorithm 2} is used, report $\|\mv g_i^{(t)}\|_1$ and $\|\mv g_i^{(t)}\|^2$ to the server\;
}
Server determines $\mathcal I^{(t)}$ and $\rho^{(t)}$\;
\ActiveParallel{
Update memory via \eqref{eq:memory update rule}\;
Transmit
$
x_{i,j}^{(t)}=\sqrt{\Myfrac{\rho^{(t)} g_{i,j}^{(t)}}{(\kappa_i \rev{\eta})}}
$ for all $j \in [d]$\;
}
\OnServer{
Receive signal $\mv y^{(t)}$\;
Compute
$r_j^{(t)}$ via \eqref{eq:NC receiver} for all $j\in[d]$\;
Update the global model via \eqref{eq:NCAirFL global update} \;
Broadcast $\mv \theta^{(t+1)}$ to all devices\;
}
Set $\mv m_i^{(t+1)}\leftarrow\mv m_i^{(t)}$ for every $i\notin\mathcal I^{(t)}$\;
$t \leftarrow t + 1$\;
}
\textbf{Output:} $\mv \theta^{(T)}$
\end{algorithm}

\section{Convergence Analysis} \label{sec:Convergence Analysis}

In this section, we analyze the convergence behavior of the proposed NCAirFL scheme for smooth and non-convex empirical risk minimization (ERM). 
\rev{To make the conditioning explicit, let $\mathcal F_0$ denote the $\sigma$-algebra induced by the initial state that determines a scheduling policy $\{\mathcal I^{(t)},\rho^{(t)}\}_{t=0}^{T-1}$ that is independent of the learning trajectory. 
For $t\ge1$, let $\mathcal F_t$ augment $\mathcal F_0$ with the mini-batch, dither, fading, and noise accrued in all previous $t-1$ rounds. Consequently, $\mv\theta^{(t)}$ and $\{\mv m_i^{(t)}\}_{i=1}^n$ are $\mathcal F_t$-measurable. 
% For local step $q$, let $\mathcal F_{t,q}$ augment $\mathcal F_t$ with the round-$t$ information revealed before the mini-batch at step $q$ is sampled, so that $\mv\theta_i^{(t,q)}$ is $\mathcal F_{t,q}$-measurable.
}
Throughout the analysis, we adopt the following standard assumptions for FL analysis~\cite{basu19qsparse,yang2021achieving}.

\begin{assumption}[$L$-smoothness] \label{assumption: L-smoothness}
For each device $i\in[n]$, the local empirical loss function $f_i(\cdot)$ is differentiable and $L$-smooth for some $L>0$, i.e., for all \(\mv x,\mv y\in\mathbb{R}^d,\)
\begin{equation}
    \left\|\nabla f_i(\mv x)-\nabla f_i(\mv y)\right\| \le L\|\mv x-\mv y\|.
    \label{eq:L-smoothness}
\end{equation}
\end{assumption}

\begin{assumption}[Bounded conditional moments of stochastic gradients] \label{assumption:bounded variance}
For every device $i\in[n]$, round $t$, and local step $q$, the stochastic gradient estimator satisfies
\begin{equation}
    \mathbb E\left[\hat\nabla f_i(\mv\theta_i^{(t,q)})\right]=\nabla f_i(\mv\theta_i^{(t,q)}),
\end{equation}
with bounded conditional variance
\begin{equation}
    \mathbb{E}\left[\left\|\hat\nabla f_i(\mv\theta_i^{(t,q)})-\nabla f_i(\mv\theta_i^{(t,q)})\right\|^2\mid\mathcal F_{t}\right] \le \sigma_l^2,
\end{equation}
and bounded conditional second moment
\begin{equation}
    \mathbb E\left[\left\|\hat\nabla f_i(\mv\theta_i^{(t,q)})\right\|^2\mid\mathcal F_t\right]\le G^2.
\end{equation}
\end{assumption}

\begin{assumption}[Bounded heterogeneity] \label{assumption:heterogeneity}
For any $\boldsymbol \theta\in \mathbb{R}^{d}$ and every device $i\in[n]$, the local gradients satisfy
\begin{equation}
    \left\|\nabla f_i(\boldsymbol \theta)-\nabla f(\boldsymbol \theta)\right\|^2 \le \sigma_g^2,
\end{equation}
where $f(\mv \theta)=(\Myfrac{1}{n})\sum_{i=1}^{n}f_i(\mv \theta)$.
\end{assumption}

\begin{assumption}[Uniform partial participation] \label{assumption:uniform participation}
% \rev{
% $\{\mathcal I^{(t)}\}_{t=0}^{T-1}$ are sampled independently at initialization. 
Each $\mathcal I^{(t)}$ is sampled independently across rounds and uniformly without replacement from the subsets of $[n]$ with cardinality $s=rn$, and the entire sequence is independent of the mini-batch, dither, fading, and noise randomness.
% }
\end{assumption}

\begin{assumption}[Lower-bounded objective] \label{assumption:lower bounded}
The global objective is bounded below: $f_*\triangleq\inf_{\mv\theta\in\mathbb R^d}f(\mv\theta)>-\infty$.
\end{assumption}

We first state the contraction property of the memory mechanism.
\begin{lemma}[Contraction] \label{lemma:contraction}
For any $i\in[n]$ and $t\in\{0,1,\ldots,T-1\}$, the preprocessing and memory update satisfy
\begin{multline}
\label{eq:contraction}
\mathbb{E}\left[\left\| \boldsymbol m_i^{(t)}+\boldsymbol \Delta_i^{(t)} - \boldsymbol \phi^{(t)}\odot \mv g_i^{(t)} \right\|^2
\,\middle|\,\mathcal F_t,\mv\Delta_i^{(t)}\right] \\
\le \left(1-\lambda \right)\left\| \boldsymbol m_i^{(t)}+\boldsymbol \Delta_i^{(t)}\right\|^2,
\end{multline}
where the remaining randomness is that of $\boldsymbol\phi^{(t)}$ and $\lambda=\min(p,1-p)$. The upper bound is minimized by $p=1/2$, for which $\lambda=1/2$.
\end{lemma}
\begin{IEEEproof}
See Appendix~\ref{appendix:sketch proof of contraction}.
\end{IEEEproof}

Under these assumptions, we study convergence in terms of the expected squared gradient norm averaged over all global rounds,
\(
(\Myfrac{1}{T})\sum_{t=0}^{T-1}\mathbb{E}[\|\nabla f(\boldsymbol\theta^{(t)})\|^2],
\)
which is a standard performance metric for smooth and non-convex optimization algorithms~\cite{ghadimi2013stochastic}. 
To this end, we define the non-coherent detection error
\begin{equation} \label{eq:NC error}
    \boldsymbol e^{(t)}=\boldsymbol r^{(t)}-\frac{1}{\rev{\eta}}\sum_{i\in \mathcal{I}^{(t)}}\boldsymbol g_i^{(t)}.
\end{equation}
By the channel assumptions and square-law detector, it satisfies the conditional zero-mean property
\begin{equation}\label{eq:conditional zero mean error}
\mathbb E\left[\mv e^{(t)}\,\middle|\,\mathcal F_t,\mathcal I^{(t)},\{\mv\Delta_i^{(t)}\}_{i=1}^n,\mv\phi^{(t)}\right]=\mv 0,
\end{equation}
Under this conditioning, only the fading and noise in round $t$ remain random in the detector.
A key step is to bound the mean-squared error (MSE) of the non-coherent detection error $\boldsymbol e^{(t)}$ in \eqref{eq:NC error}.

\begin{lemma}[MSE of non-coherent detection] \label{lemma:MSE of NC detection}
\rev{Under Assumption~\ref{assumption:bounded variance}, if $\rho^{(t)}$ is the largest possible common scaling coefficient satisfying \eqref{eq:power constraint}, then the MSE of the non-coherent detector satisfies}
\begin{multline}\label{eq:MSE bound}
    \mathbb{E}\!\left[\|\mv e^{(t)}\|^2\,\middle|\,\rev{\mathcal F_0}\right] \le \\
    M_h s\tilde{G}^2 + \frac{\tilde{G}^2}{\rho_{\min}^2} + 2s(s-1)\tilde{G}^2 +\frac{2 s \tilde{G}^2}{\rho_{\min}},
\end{multline}
\rev{where the conditional expectation is taken over the learning and communication induced randomness. 
% Since the right-hand side is deterministic, the same bound holds after taking total expectation.
} In addition,
\(
\tilde{G}^2=2\left( \Myfrac{4(1-\lambda^2)}{\lambda^2} +1 \right)Q^2G^2
\)
with
\(
\lambda=\min(p,1-p),
\)
and
\(
\rho_{\min}=\min_{i\in[n]}\Myfrac{P_i\kappa_i}{\sigma^2}.
\)
\end{lemma}

\begin{IEEEproof}
See Appendix~\ref{appendix:sketch proof of MSE}.
\end{IEEEproof}

The bound in~\eqref{eq:MSE bound} reveals the compositional structure of the non-coherent detection error in terms of the second-moment bound $\tilde G^2$.
The first term, \(M_hs\tilde{G}^2\), captures the effect of the fourth moment of the fading distribution and is nonnegative.
The second and fourth terms stem from noise--noise and signal--noise products and decay as \(1/\rho_{\min}^2\) and \(s/\rho_{\min}\), respectively. 
The third term arises from cross products among different devices' signals and scales as $s(s-1)$, quantifying the additional interference created when more devices participate in non-coherent aggregation.

We are now ready to present the main convergence result for NCAirFL under a constant learning rate.

\begin{proposition}[Convergence] \label{proposition:convergence}
Suppose that Assumptions~\ref{assumption: L-smoothness}--\ref{assumption:lower bounded} hold.
Let $\{\boldsymbol \theta^{(t)}\}$ be generated by NCAirFL over $T$ communication rounds with constant learning rate \rev{$\eta$} satisfying $\eta \in(0, 1/(\sqrt{240}QL)]$. Then, averaged over the randomness of mini-batch sampling, dithering, device selection, small-scale fading, and channel noise, NCAirFL satisfies
\begin{multline} \label{eq:convergence bound}
\frac{1}{T}\sum_{t=0}^{T-1}\mathbb{E}\left[\|\nabla f(\boldsymbol \theta^{(t)})\|^2\right]
\le
\underbrace{\frac{8(f_0-f_*)}{T \eta Q}}_{\text{Initialization error}}
\\
+
\underbrace{\frac{4\eta LG_e^2}{Qs^2}}_{\text{Non-coherent detection error}}
+
\underbrace{4\eta L Q G^2
+40\eta^2 Q L^2\left(\sigma_l^2+6Q\sigma_g^2\right)}_{\text{Stochastic-gradient and local-update error}}
\\ 
+
\underbrace{\frac{48 \eta^2 Q^2 L^2(1-\lambda^2)G^2}{r^2\lambda^2}}_{\text{Contraction error}},
\end{multline}
where \(f_0=f(\mv \theta^{(0)})\), $f_*$ is defined in Assumption~\ref{assumption:lower bounded}, 
and
\(G_e^2\) denotes the deterministic upper bound on $\mathbb{E}[\|\mv e^{(t)}\|^2]$ given by the right hand side (RHS) of \eqref{eq:MSE bound}.
\end{proposition}

\begin{IEEEproof}
See Appendix~\ref{appendix:sketch proof of convergence}.
\end{IEEEproof}

The bound in \eqref{eq:convergence bound} reveals how different sources of error affect the learning performance of NCAirFL. The first term is the initialization error, which decays as the number of communication rounds $T$ grows. 
The second term captures the impact of non-coherent aggregation, which increases with the upper bound on MSE $\mathbb{E}\|\mv e^{(t)}\|^2$ and decreases with the worst device's average (received) SNR as expected. 
The third term collects the errors due to bounded stochastic gradients and the drift generated by multiple local updates. 
The last term arises from the accumulated memory and depends on the dithering parameter $\lambda$.

By setting the learning rate as $\eta=\mathcal{O}(1/\sqrt{T})$, the bound in \eqref{eq:convergence bound} implies convergence to a stationary point at rate $\mathcal{O}(1/\sqrt{T})$. This matches the standard order-wise convergence rate of FedAvg for smooth non-convex objectives~\cite[Theorem~1]{yang2021achieving}, despite NCAirFL operating without instantaneous CSI.

\section{Joint Optimization of Device Selection and Power Control} \label{sec:Optimization}

Although over-the-air aggregation uses the same wireless resource regardless of the active-device count, activating every device is not always optimal here. 
For example, in \eqref{eq:MSE bound}, the non-coherent inter-device term grows as $2s(s-1)\tilde G^2$. Moreover, for any selected set, the feasible common scaling is limited by its most power-constrained device.
Partial participation can therefore reduce the quadratic cross-term and remove a power bottleneck, at the cost of discarding useful innovations. 
Centralized selection coordinates this trade-off between aggregate learning utility and a common power scaling. Independent local decisions cannot in general identify the common bottleneck.

In this section, we study joint optimization of device selection and power control to improve on the convergence performance of NCAirFL. 
To this end, we first derive a lower bound on the expected one-round forward reduction for any prescribed active device set.
We then formulate joint device selection and power control using this bound. Because the required moments are unavailable in closed form, we replace them with causal estimates and solve the resulting surrogate problem optimally. The resulting policy is evaluated in Section~\ref{sec:Numerical Results}.

\subsection{Optimization Problem  Formulation}

\begin{proposition}[One-round forward reduction] \label{proposition:single-round reduction}
\rev{Suppose that Assumptions~\ref{assumption: L-smoothness}--\ref{assumption:heterogeneity} hold and that $\mathcal I^{(t)}\subseteq[n]$ and $\rho^{(t)}>0$ are $\mathcal F_0$-measurable satisfying \eqref{eq:power constraint}. 
Letting $\eta\in(0,1/(\sqrt{2}QL)]$, then NCAirFL satisfies}
\begin{multline} \label{eq:single-round reduction}
    \mathbb E\left[f(\boldsymbol{\theta}^{(t)}) - f(\boldsymbol{\theta}^{(t+1)})\,\middle|\,\rev{\mathcal F_0}\right] \\
    \ge \underbrace{\left(\frac{2-L \eta Q}{2s\eta Q}\right)\sum_{i\in\mathcal{I}^{(t)}}\mathbb{E}\left[\| \boldsymbol g_i^{(t)} \|^2\,\middle|\,\rev{\mathcal F_0}\right]}_{\text{Utility}} \\
    - \underbrace{\frac{1}{s}\sum_{i\in\mathcal{I}^{(t)}}C_1\sqrt{\mathbb{E}\left[\| \boldsymbol g_i^{(t)} \|^2\,\middle|\,\rev{\mathcal F_0}\right]}}_{\text{SGD error and data heterogeneity}} \\
    -\underbrace{\frac{\eta^2 L}{2s^2}\left(
    (M_h+2s-2)s\tilde{G}^2+\frac{d\sigma^4}{(\rho^{(t)})^2}
    +\frac{2s\sqrt{d}\sigma^2}{\rho^{(t)}}\tilde{G}\right)}_{\text{Effective noise}},
\end{multline}
where
\(
C_1=
\sqrt{5L^2Q\eta^2\!\left(\sigma_l^2+6Q\sigma_g^2\right)
+30L^2Q^2\eta^2G^2}
+\sigma_g+\Myfrac{\sigma_l}{\sqrt Q}
+\Myfrac{4G\sqrt{1-\lambda^2}}{\lambda}.
\)
% with $\lambda=\min(p,1-p)$, and
% \(
% \tilde{G}^2=2\left(\frac{4(1-\lambda^2)}{\lambda^2}+1\right)Q^2G^2.
% \)
\end{proposition}

\begin{IEEEproof}
See Appendix~\ref{appendix:sketch proof of single-round reduction}.
\end{IEEEproof}

Proposition~\ref{proposition:single-round reduction} characterizes the expected one-round decrease in the objective function given each prescribed device scheduling. 
First, the utility term favors devices with larger \rev{$\mathbb E[\|\mv g_i^{(t)}\|^2\mid\mathcal F_0]$}, reflecting larger potential descent contributions.
Second, the SGD-error and data-heterogeneity term accounts for stochastic-gradient noise, data heterogeneity, and contraction error. 
Third, the effective-noise term captures the loss in one-round descent caused by non-coherent aggregation through the fading moment $M_h$ and the power scaling coefficient $\rho^{(t)}$. 
These terms motivate joint optimization of device selection \(\mathcal I^{(t)}\) and power control \(\rho^{(t)}\) in each communication round.
Fixing round $t$ and suppressing time superscripts on the decision variables, we formulate
\begin{subequations} \label{eq:selection optimization}
\begin{align}
\mathrm{(P1)}: 
\operatorname*{\mathtt{Maximize}}_{\mathcal{I}, \rho} \  &
U^{(t)}(\mathcal{I}) - \frac{\eta^2 L}{2s^2}\left(\frac{2s\sqrt{d}\sigma^2}{\rho}\tilde{G}+\frac{d\sigma^4}{\rho^2}\right) \nonumber \\
\mathtt{Subject \ to} \quad  &
\mathcal{I}\subseteq [n],\ |\mathcal{I}|=s, \quad \rho>0,  \label{eq:constraint of selection} \\
&
\frac{\rho}{\kappa_i\eta}\mathbb{E}\!\left[\|\boldsymbol g_i^{(t)}\|_1\,\middle|\,\rev{\mathcal F_0}\right] \le dP_i,
\quad \forall i\in\mathcal I. \label{eq:constraint of power in selection} 
\end{align}
\end{subequations}    
Here, $U^{(t)}(\mathcal I)$ denotes the sum of the utility term and the negative SGD-error/data-heterogeneity term in \eqref{eq:single-round reduction}; constraint~\eqref{eq:constraint of power in selection} enforces the average-power constraint~\eqref{eq:power constraint}. 
% \hf{The $\ell_1$ norm in \eqref{eq:constraint of power in selection} follows directly from $g_{i,j}^{(t)}\ge0$: by \eqref{eq:non-coherent transmitted signal}, $\|\mv x_i^{(t)}\|^2=(\rho/(\kappa_i\eta))\sum_j g_{i,j}^{(t)}=(\rho/(\kappa_i\eta))\|\mv g_i^{(t)}\|_1$.}

\subsection{Proposed solution to (P1)}
Problem \(\mathrm{(P1)}\) depends on the moments \rev{$\mathbb E[\|\mv g_i^{(t)}\|^2\mid\mathcal F_0]$ and $\mathbb E[\|\mv g_i^{(t)}\|_1\mid\mathcal F_0]$}, which are unavailable in closed form. We therefore replace the corresponding moments by causal exponential moving averages (EMAs) with the bias correction used in Adam~\cite{kingma2014adam}. In every round, all devices  perform the local update in \eqref{eq:local updates}, construct the candidate $\mv g_i^{(t)}$ from the current innovation and memory, and send the two real scalars $\|\mv g_i^{(t)}\|^2$ and $\|\mv g_i^{(t)}\|_1$ to the server over a reliable low-rate control channel before selection. The resulting control overhead is $2n$ real scalars per round, which tends to be negligible since $d\gg 2n$ typically. Only the selected devices commit the memory update in \eqref{eq:memory update rule} and transmit the model payload.

Initialize $u_i^{(-1)}=v_i^{(-1)}=0$. Because each device \(i\) uploads statistics \(\|\boldsymbol g_i^{(t)}\|^2\) and \(\|\boldsymbol g_i^{(t)}\|_1\) in every round \(t\), the server updates
\begin{gather}
v_i^{(t)} = \beta v_i^{(t-1)} + (1-\beta)\|\boldsymbol g_i^{(t)}\|^2, \label{eq:approximate l2 norm of expected g} \\
u_i^{(t)} = \beta u_i^{(t-1)} + (1-\beta)\|\boldsymbol g_i^{(t)}\|_1,
\end{gather}
where $\beta\in[0,1)$ is the forgetting factor. We call \cite{kingma2014adam}
\begin{equation}
\widehat v_i^{(t)}=\frac{v_i^{(t)}}{1-\beta^{t+1}}, \qquad \widehat u_i^{(t)}=\frac{u_i^{(t)}}{1-\beta^{t+1}}
\end{equation}
the \emph{normalized EMA statistics}, which serve as causal \rev{surrogates for the corresponding second and first-order moments otherwise inaccessible}, respectively. Since every device refreshes these statistics in every round, an unselected device's EMA does not decay merely because it was inactive.
% As an optional fairness safeguard, one active position can be reserved for the longest-idle device once a prescribed age threshold is reached, with Algorithm~\ref{alg:Algorithm 2} applied to the remaining positions.

Accordingly, we also define the approximate utility with penalty by SGD error and data heterogeneity of device $i \in [n]$ as
\begin{equation}
\widehat U_i^{(t)}=\left(\frac{2-L\eta Q}{2s\eta Q}\right)\widehat v_i^{(t)}-\frac{C_1}{s}\sqrt{\widehat v_i^{(t)}}.
\end{equation}
Replacing $U^{(t)}(\mathcal I)$ by $\sum_{i\in\mathcal I}\widehat U_i^{(t)}$ and $\mathbb E[\|\mv g_i^{(t)}\|_1]$ by $\widehat u_i^{(t)}$ thus yields the surrogate problem \(\mathrm{(P2)}\), where we omit the round index $t$ below for notational simplicity.
\begin{subequations} \label{eq:approximate selection optimization}
\begin{align}
\hspace{-0.1in}\mathrm{(P2)}: \ 
\operatorname*{\mathtt{Maximize}}_{\mathcal{I}, \rho} \  &
\sum_{i\in\mathcal{I}} \widehat U_i - \frac{\eta^2 L}{2s^2}\left(\frac{2s\sqrt{d}\sigma^2}{\rho}\tilde{G}+\frac{d\sigma^4}{\rho^2}\right)  \nonumber \\
\mathtt{Subject \ to} \  &
\mathcal{I}\subseteq [n],\ |\mathcal{I}|=s, \\
&
\frac{\rho}{\kappa_i\eta}\widehat u_i \le dP_i,\quad \rho>0,\quad \forall i\in\mathcal{I}.
\end{align}
\end{subequations}

Problem \(\mathrm{(P2)}\) remains combinatorial because of the set variable $\mathcal I$. Algorithm~\ref{alg:Algorithm 2} nevertheless solves it exactly by separating the common power bottleneck from the additive device utilities.

First, we observe that given any fixed active device set $\bar{\mathcal{I}}$, the optimal power scaling coefficient $\rho^{*}(\bar{\mathcal{I}})$ can be obtained in a closed form.
Specifically, since the objective in problem \(\mathrm{(P2)}\) strictly increases with $\rho>0$, the optimal $\rho$ admits the largest feasible value, i.e.,
\begin{equation}\label{eq:optimal rho}
    \rho^{*}(\bar{\mathcal{I}})=\min_{i\in\bar{\mathcal{I}}}\frac{d\kappa_i \eta P_i}{\widehat u_i}=\min_{i\in \bar{\mathcal{I}}} b_i.
\end{equation}
Next, we optimize device selection by substituting $\rho^{*}(\mathcal{I})$ into problem \(\mathrm{(P2)}\) using \eqref{eq:optimal rho}, which yields an equivalent problem:
\begin{align}
\mathrm{(P2')}:
\operatorname*{\mathtt{Maximize}}_{\mathcal{I}} \  &
\sum_{i\in\mathcal{I}} \widehat U_i-\frac{\eta^2 L}{2s^2}\left(\frac{2s\sqrt{d}\sigma^2}{\rho^{*}(\mathcal{I})}\tilde{G}+\frac{d\sigma^4}{(\rho^{*}(\mathcal{I}))^2}\right) \nonumber \\
\mathtt{Subject \ to} \  &
\mathcal{I}\subseteq [n],\ |\mathcal{I}|=s.
\label{eq:subproblem It}
\end{align}

Note that Algorithm~\ref{alg:Algorithm 2} sorts these values, treats each device in turn as the candidate bottleneck, selects the remaining $s-1$ devices with the largest utilities compatible with that bottleneck, and evaluates the resulting objective. Enumerating all possible bottlenecks yields the global optimum without exhaustive subset search.

\begin{proposition}[Optimal Solution to (P2)]\label{proposition:optimal solution P1}
     Suppose that $\widehat u_i>0$ for every $i\in[n]$. Algorithm~\ref{alg:Algorithm 2} yields the jointly optimal active device set \(\mathcal{I}^{*}\) and power control factor \(\rho^*\).
\end{proposition}
\begin{IEEEproof}
To solve problem \(\mathrm{(P2')}\), define
\(
b_i=\Myfrac{d\kappa_i \eta P_i}{\widehat u_i},
\)
which, by \eqref{eq:optimal rho}, gives
\(
\rho^{*}(\mathcal{I})=\min_{i\in\mathcal{I}} b_i.
\)
% Hence, for any feasible set $\mathcal{I}$, the optimal power scaling factor is determined by the selected device with the smallest $b_i$.
Let $\pi$ be a permutation of $[n]$ such that
\(
b_{\pi(1)} \ge b_{\pi(2)} \ge \cdots \ge b_{\pi(n)}.
\)
Then, fixing an index $k\in\{s,s+1,\ldots,n\}$, we consider the case in which the bottleneck device is $\pi(k)$, i.e., $\rho^{*}(\mathcal{I})=b_{\pi(k)}$. This case is feasible if and only if $\pi(k)\in\mathcal{I}$ and $\mathcal{I}\subseteq\{\pi(1),\ldots,\pi(k)\}$. 
Therefore, for a fixed bottleneck $\pi(k)$, the optimal set includes $\pi(k)$ and the $s-1$ devices with the largest $\widehat U_i$ among $\{\pi(1),\ldots,\pi(k-1)\}$. Enumerating $k\in\{s,s+1,\ldots,n\}$ then yields a globally optimal solution to \(\mathrm{(P2)}\).
\end{IEEEproof}
% The procedure is summarized in Algorithm~\ref{alg:Algorithm 2}. 

\begin{algorithm}[t]
\SetKwInOut{Input}{Input}
\SetKwInOut{Output}{Output}
\SetKwBlock{ForLoop}{for $k=s,s+1,\ldots,n$ do}{end}
\caption{Optimal Solution to Problem $\mathrm{(P2)}$} \label{alg:Algorithm 2}
\Input{$\{\widehat U_i\}_{i=1}^{n}$, $\{\widehat u_i\}_{i=1}^{n}$, $\{P_i\}_{i=1}^{n}$, $\{\kappa_i\}_{i=1}^{n}$,  $\tilde{G}$, \rev{$s$}, $\eta$, $\sigma$, $L$, and $d$}
Compute $b_i=\Myfrac{d\kappa_i \eta P_i}{\widehat u_i}$ for all $i\in[n]$, and sort them such that $b_{\pi(1)}\ge b_{\pi(2)}\ge \cdots \ge b_{\pi(n)}$\;
Initialize $y^{*}=-\infty$\;
\ForLoop{
Let $\mathcal{T}_k=\{\pi(1),\ldots,\pi(k-1)\}$\;
% Choose $\mathcal{J}_k\subseteq\mathcal{T}_k$ as the set of the $s-1$ devices with the largest utilities $\{\widehat U_i:i\in\mathcal{T}_k\}$\;
Choose \( \mathcal{J}_k = \operatorname*{argmax}_{\mathcal{J}\subseteq\mathcal{T}_k,|\mathcal J|=s-1} \sum_{i \in \mathcal{J}}\widehat U_i \)\;
Set $\mathcal{I}_k=\mathcal{J}_k\cup\{\pi(k)\}$ and $\rho_k=b_{\pi(k)}$\;
Compute
\(
y_k=\sum_{i\in\mathcal{I}_k}\widehat U_i-\frac{\eta^2L}{2s^2}(\frac{\rev{2s}\sqrt{d}\sigma^2}{\rho_k}\tilde{G}+\frac{d\sigma^4}{(\rho_k)^2})
\)\;
Update $(y^{*},\mathcal{I}^{*})\leftarrow (y_k,\mathcal{I}_k)$ if $y_k>y^{*}$\;
}
\Output{$\mathcal{I}^{*}$, and \(\rho^{*}=\min_{i\in\mathcal{I}^{*}} \Myfrac{d\kappa_i \eta P_i}{\widehat u_i}\)}
\end{algorithm}

\emph{Complexity of Algorithm \ref{alg:Algorithm 2}:} Sorting $\{b_i\}_{i=1}^n$ requires $\mathcal{O}(n\log n)$ operations. For each $k\in\{s,\ldots,n\}$, selecting the $s-1$ largest utilities among the first $k-1$ devices and evaluating $y_k$ can be implemented in $\mathcal{O}(n)$ time. Therefore, the overall complexity of Algorithm~\ref{alg:Algorithm 2} is $\mathcal{O}(n^2+n\log n)$, which is much lower than exhaustive search over \(\binom{n}{s}\) feasible subsets, which admits an exponential complexity order \( \mathcal{O}(2^{nH(r)})\), where $H(r) = -r \log_2 r - (1-r) \log_2(1-r)$ \cite{cover1991elements}.

\begin{remark}
% Following a common convention in partial-participation FL~\cite{mcmahan2017communication,yue2022efficient,yang2021achieving}, problems~\(\mathrm{(P1)}\) and~\(\mathrm{(P2)}\) treat the participation ratio $r$ as a prescribed system parameter. 
It's worth noting that the bound on single-round reduction in Proposition~\ref{proposition:single-round reduction} also permits joint optimization of $r$, $\mathcal I^{(t)}$, and $\rho^{(t)}$.
Specifically, by letting $m=s$ denote the active device count and enumerating candidate bottleneck devices over $m$, the remaining $m-1$ devices can be chosen based on their gradient-descent contributions, i.e., \(\widehat U_i\). By reusing the sorted gradient-descent contributions and their cumulative sums, we obtain the optimal solution with the same complexity $\mathcal{O}(n^2+n\log n)$ as in Algorithm~\ref{alg:Algorithm 2}. 
We leave this joint formulation as a straightforward extension and focus on a fixed $r$ in this work.
\end{remark}

\begin{remark}
Proposition~\ref{proposition:convergence} assumes uniform sampling without replacement, under which active-device averaging yields an unbiased estimate of the global-average update. 
By contrast, Algorithm~\ref{alg:Algorithm 2} selects devices deterministically based on state-dependent statistics, i.e., \(\widehat u_i\) and \(\widehat v_i\), and power constraints \eqref{eq:constraint of power in selection}, systematically favoring devices with larger gradient-descent contributions and favorable channel conditions. 
A similar approach is adopted in \cite{yue2022efficient,nguyen2021fast}, where non-uniform device selection maximizes a lower bound of the one-round loss reduction while leaving global convergence under such device selection policy not addressed. 
On the other hand, probabilistic scheduling methods \cite{ren2020scheduling,sun2024channel} maintain unbiased aggregation via probability-aware reweighting, enabling end-to-end convergence analysis under non-uniform participation. 
Extending the convergence of NCAirFL to support a state-dependent device-selection policy is left for future work.
\end{remark}

% \begin{remark}
%     \rev{Proposition~\ref{proposition:single-round reduction} provides the analytical basis for jointly optimizing the active-device set and the common power-scaling factor $\rho^{(t)}$ by balancing learning utility against stochastic-gradient, data-heterogeneity, and communication penalties. Algorithm~\ref{alg:Algorithm 2} replaces the unavailable expectations with state-dependent EMA statistics and optimizes the resulting surrogate. Since the optimal active device set depends on the realized learning history, however, Proposition~\ref{proposition:single-round reduction} does not provide a corresponding conditional lower bound for that realized selection.}
% \end{remark}

\section{Numerical Results} \label{sec:Numerical Results}
In this section, we evaluate NCAirFL with $n=20$ devices on MNIST and CIFAR-10. We examine how closely NCAirFL approximates communication-ideal FedAvg and quantify the gain from the proposed joint device-selection and power-control strategy.

\subsection{Experimental Settings} \label{subsec:Experimental Settings}
We consider two standard image-classification benchmarks, namely MNIST and CIFAR-10. MNIST contains $60{,}000$ training samples and $10{,}000$ test samples, where each sample is a grayscale image of size $28\times 28$. CIFAR-10 contains $50{,}000$ training images and $10{,}000$ test images from $10$ classes, with $6{,}000$ images per class and image size $32\times 32$~\cite{lecun1998mnist, krizhevsky2009cifar10}.

For MNIST, we consider a non-i.i.d. data split generated via a Dirichlet distribution with concentration parameter $\varrho=0.1$~\cite{yurochkin2019bayesian}.\footnote{For each class $k\in[K]$, a probability vector $\mv q_k\sim \mathrm{Dir}(\varrho \mv 1_n)$ is sampled, where the $j$-th entry $q_{k,j}$ specifies the fraction of class-$k$ samples assigned to device $j$. Hence, smaller $\varrho$ corresponds to more severe label-distribution skew.} The model is a two-layer fully connected network with a flattened $28\times 28$ input, one hidden layer of $100$ ReLU units, and a softmax output layer, for a total of $d=79{,}510$ trainable parameters.

For CIFAR-10, we focus on the i.i.d. setting, in which the local dataset $\mathcal{D}_i$ at device $i\in[n]$ is sampled uniformly at random without replacement from the training set. The model is ResNet-18, with $d=11{,}181{,}642$ trainable parameters~\cite{he2016deep}.

For the wireless channel, we adopt the path-loss model $\kappa_i=\Myfrac{c^2}{(4\pi f_c R_i)^2}$ for $i\in[n]$, where $c$ is the speed of light, $f_c=2.4$\,GHz is the carrier frequency, and \rev{deploy devices uniformly over the area of a disk of radius $1000$\,m centered at the server, yielding the distance $R_i\sim 1000\sqrt{\mathcal{U}(0,1)}$.}
Small-scale fading is Rayleigh. Unless stated otherwise, we use learning rates $\eta=0.001$ for MNIST and $\eta=0.05$ for CIFAR-10, mini-batch size $|\mathcal{B}_i^{(t,q)}|=64$, $Q=5$ local SGD steps, participation ratio $r=0.5$, dithering parameter $p=1/2$, and EMA forgetting factor $\beta=0.1$. 
The small value of $\beta$ makes the selection statistics emphasize the current innovation.
In solving the approximate selection problem in~\eqref{eq:approximate selection optimization}, we set $G=L=\sigma_l=\sigma_g^2=0.1$. 
The transmit-power constraint is $P_i=3$\,dBm  for all $i\in[n]$, and the noise power is obtained from power spectral density $N_0=-173$\,dBm/Hz over bandwidth $20$\,MHz, yielding $\sigma^2\approx -100$\,dBm. Under this configuration, the maximum received SNR is approximately $13$\,dB. We use SGD as the optimizer, and all reported curves are averaged over $50$ Monte Carlo trials.

We compare NCAirFL against the following baselines: (\emph{i}) FedAvg with ideal communication~\cite{mcmahan2017communication}; (\emph{ii}) FedAvg with optimized device selection, denoted by \emph{FedAvg(Opt.)}, obtained by solving
\begin{align}
    \operatorname*{\mathtt{Maximize}}_{\mathcal{I}^{(t)}} \quad & a_1 \sum_{i\in\mathcal{I}^{(t)}}\mathbb{E} \left\| \boldsymbol \Delta_i^{(t)} \right\|^2  - \frac{a_2}{s}\sum_{i\in\mathcal{I}^{(t)}}\sqrt{\mathbb{E} \left\| \boldsymbol \Delta_i^{(t)} \right\|^2} \nonumber \\
    \mathtt{Subject \ to} \quad & \mathcal{I}^{(t)}\subseteq [n],\ |\mathcal{I}^{(t)}|=s,
\end{align}
where $a_1$ and $a_2$ are tuned constants and $\mathbb{E}\|\boldsymbol \Delta_i^{(t)}\|^2$ is approximated as in~\eqref{eq:approximate l2 norm of expected g}; (\emph{iii}) coherent AirFL with truncated channel inversion, referred to as \emph{CAirFL}~\cite{zhu2019broadband}; and (\emph{iv}) \emph{AirFL-Mem}, which augments truncated-CI AirFL with long-term memory to mitigate deep-fading errors~\cite{wen2024AirFL-Mem}. The hyperparameters of CAirFL and AirFL-Mem follow~\cite{wen2024AirFL-Mem}. Table~\ref{table:convergence rate comparisons} summarizes the main theoretical differences among the compared methods. We use \emph{NCAirFL(Opt.)} to denote the proposed method combined with the optimized device-selection rule.
For coherent baselines, in a typical 5G NR configuration employing DM-RS configuration type 1, the CSI estimation overhead is approximately \(14.3\%\) \cite{3GPP}, which is eliminated by NCAirFL.

\begin{table}[t]
\centering
\caption{Properties of the considered FL schemes} \label{table:convergence rate comparisons}
\begin{tabular}{lccc}
\hline
\textbf{Schemes}             &  \textbf{\begin{tabular}[c]{@{}c@{}}Convergence \\ rate\end{tabular}}         & \textbf{Error floor} &  \textbf{\begin{tabular}[c]{@{}c@{}}CSI \\ requirements\end{tabular}} \\ \hline
FedAvg~\cite{mcmahan2017communication} & $\mathcal{O}(1/\sqrt{T})$         & No  & N/A                                                \\
CAirFL~\cite{zhu2019broadband}         & $\mathcal{O}(1/\sqrt{T}+\text{bias})$         & Yes    & \textcolor{black}{CSIT}                           \\
AirFL-Mem~\cite{wen2024AirFL-Mem}                    & $\mathcal{O}(1/\sqrt{T})$         & No      & \textcolor{black}{CSIT}                           \\
\textbf{NCAirFL}             & $\mathcal{O}(1/\sqrt{T})$ & \textbf{No}     & \textcolor{black}{\textbf{Large-scale CSI}}                                \\ \hline
\end{tabular}
\end{table}

\subsection{Training Performance}
\begin{figure}[ht]
    \centering
    \includegraphics[width=3.2in]{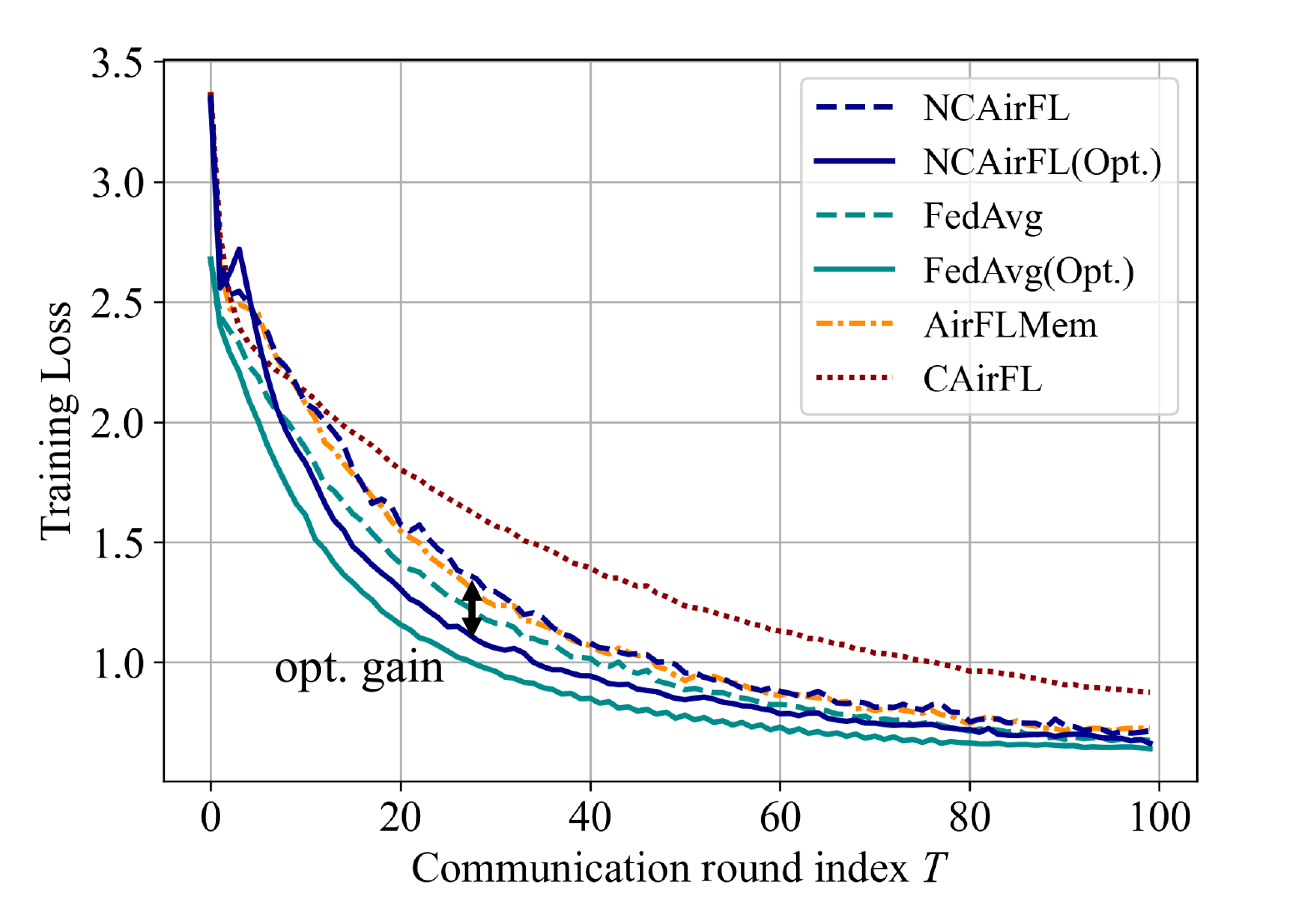}
    \vspace{-0.1in}
    \caption{Training loss versus communication round index $T$ on MNIST.}
    \label{fig:loss_vs_T_mnist}
\end{figure}

\begin{figure}[ht]
    \centering
    \includegraphics[width=3.2in]{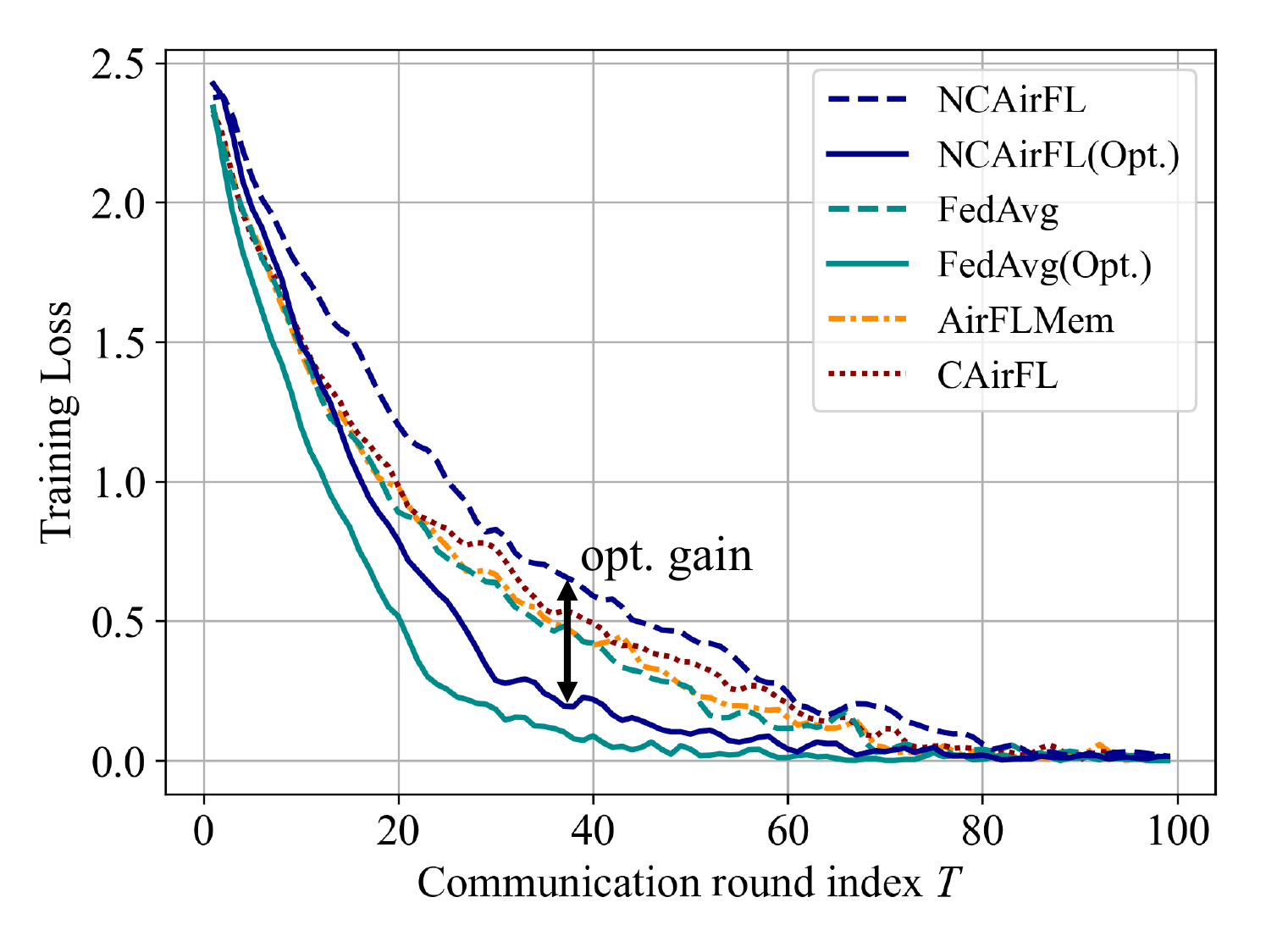}
    \vspace{-0.1in}
    \caption{Training loss versus communication round index $T$ on CIFAR-10.}
    \label{fig:loss_vs_T_cifar}
\end{figure}

Figs.~\ref{fig:loss_vs_T_mnist} and~\ref{fig:loss_vs_T_cifar} report the training loss as a function of the communication round index $T$ on MNIST and CIFAR-10, respectively. 
First, on both datasets, NCAirFL and NCAirFL(Opt.) closely track their ideal-communication counterparts, FedAvg and FedAvg(Opt.), respectively. This behavior is consistent with Proposition~\ref{proposition:convergence}, and indicates that the proposed CSI semi-free aggregation mechanism incurs only a limited penalty in training dynamics.

Second, the proposed method does not exhibit the error-floor behavior observed for CAirFL. This is particularly evident in Fig.~\ref{fig:loss_vs_T_mnist}, where CAirFL settles at a noticeably larger loss, whereas NCAirFL continues to decrease and remains close to both FedAvg and AirFL-Mem. This empirical trend is in line with the theoretical comparison in Table~\ref{table:convergence rate comparisons}.

Third, optimized device selection consistently accelerates convergence. On MNIST, the gap between NCAirFL and NCAirFL(Opt.) is already visible in the transient regime. The effect is even more pronounced on CIFAR-10, where Fig.~\ref{fig:loss_vs_T_cifar} shows that NCAirFL(Opt.) reaches a near-zero training loss within roughly $50$ rounds, while the vanilla NCAirFL still exhibits a clearly non-negligible residual loss at the same communication budget. This observation is consistent with the single-round reduction analysis in Proposition~\ref{proposition:single-round reduction}.

Fourth, AirFL-Mem outperforms vanilla NCAirFL because coherent aggregation exploits CSI. NCAirFL(Opt.), however, can outperform AirFL-Mem because optimized device selection and power control compensate for part of the non-coherent aggregation error. Thus, the end-to-end learning gain from scheduling and power control can outperform the communication advantage of the unoptimized coherent baseline in the considered settings.


\begin{figure}[ht]
    \centering
    \includegraphics[width=3.2in]{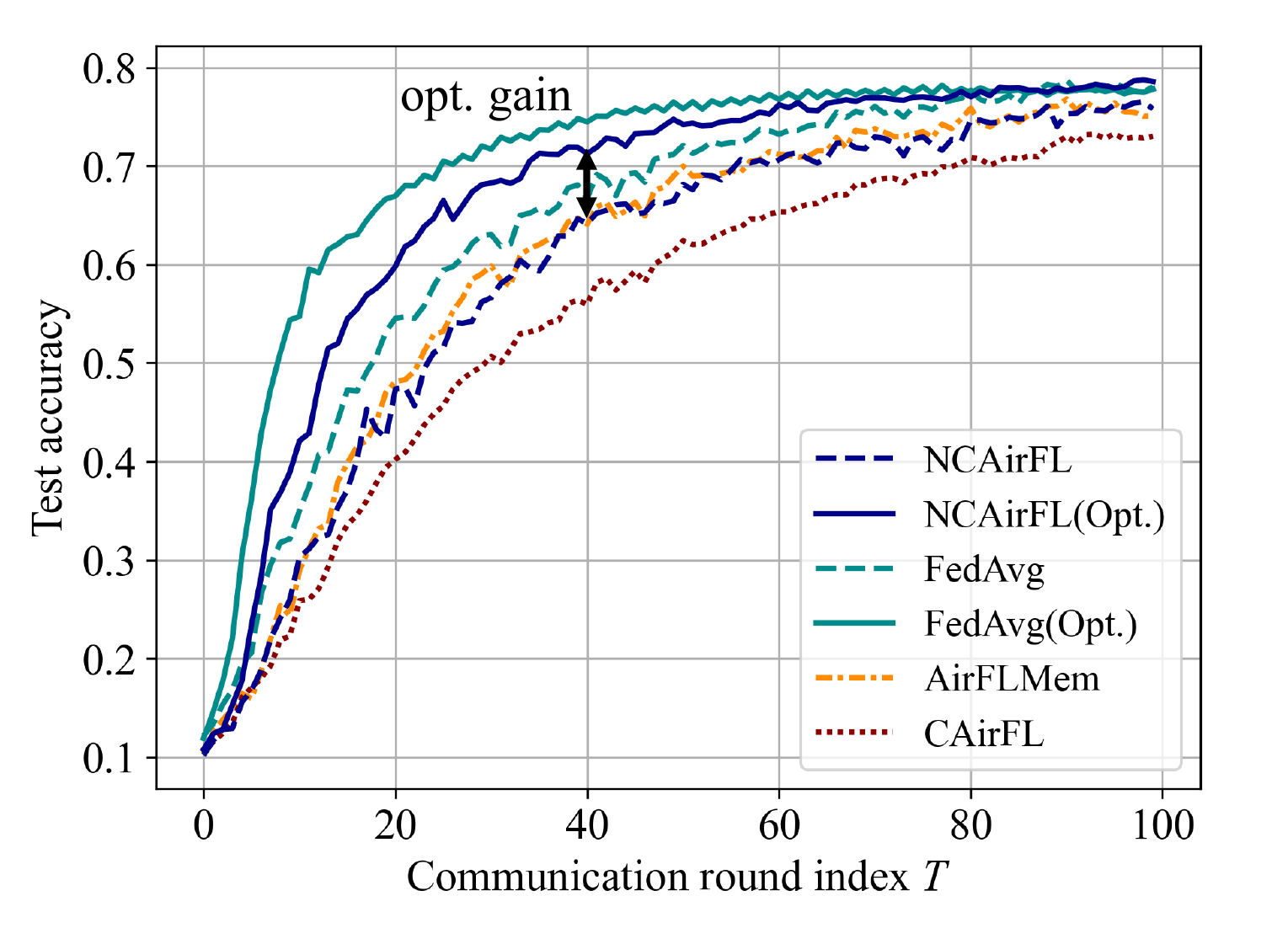}
    \vspace{-0.1in}
    \caption{Test accuracy versus communication round index $T$ on MNIST.}
    \label{fig:acc_vs_T_mnist}
\end{figure}

\begin{figure}[ht]
    \centering
    \includegraphics[width=3.2in]{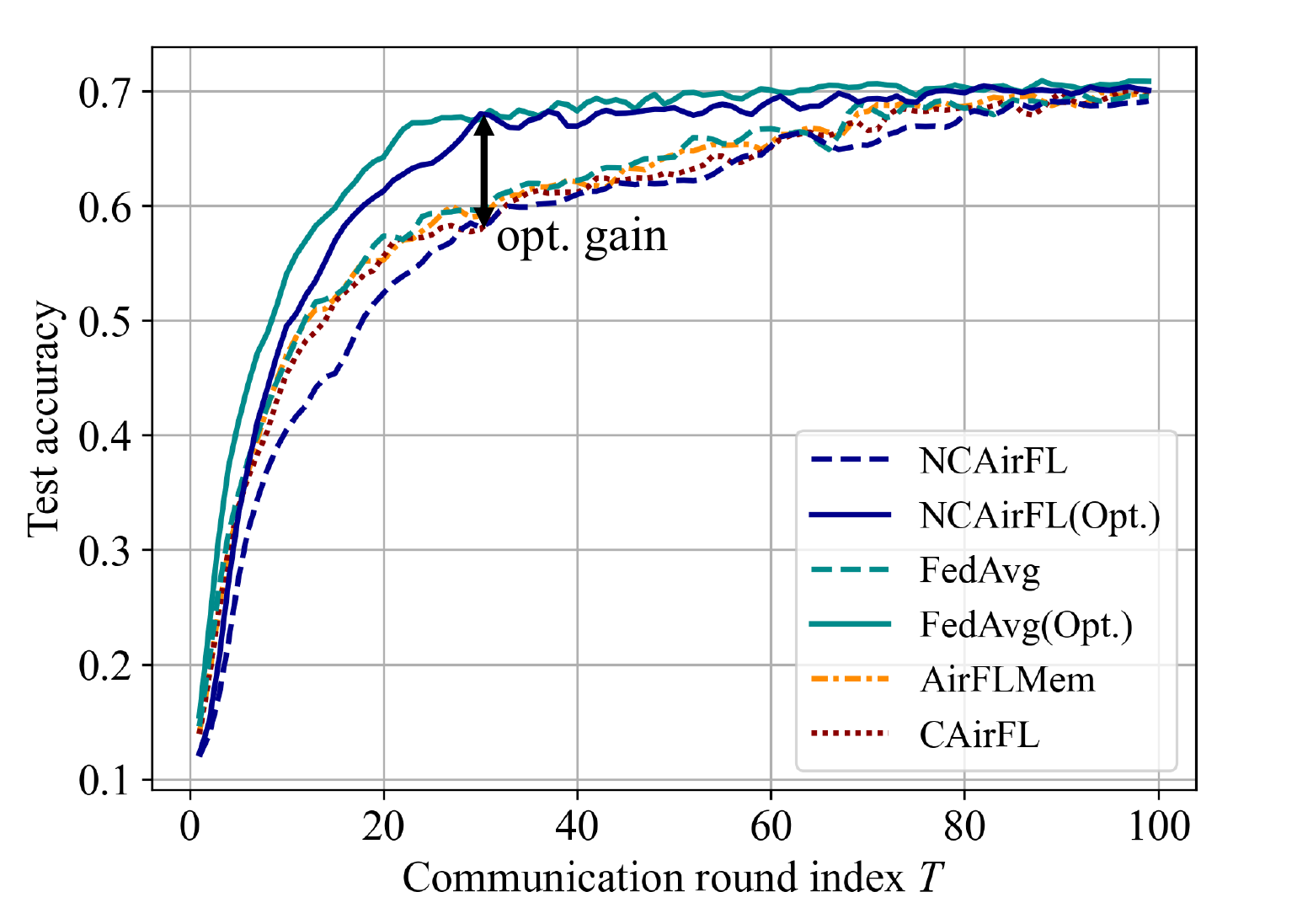}
    \vspace{-0.1in}
    \caption{Test accuracy versus communication round index $T$ on CIFAR-10.}
    \label{fig:acc_vs_T_cifar}
\end{figure}

% \subsection{Test Performance}
We next evaluate the test accuracy, which is the practically relevant learning metric. Figs.~\ref{fig:acc_vs_T_mnist} and~\ref{fig:acc_vs_T_cifar} show the test accuracy versus the communication round index $T$ on MNIST and CIFAR-10, respectively. The results mirror the trends observed for training loss.

In particular, NCAirFL attains test accuracy that is very close to FedAvg, despite operating without instantaneous CSI and over a fading wireless channel. Moreover, the optimized version NCAirFL(Opt.) improves the early- and mid-stage learning behavior on both datasets, achieving higher accuracy than vanilla NCAirFL and coherent baselines at the same communication budget. The advantage is again more visible on CIFAR-10, where the more challenging task makes the value of selecting informative and communication-efficient devices particularly clear. Overall, the results confirm that the proposed non-coherent aggregation scheme preserves not only convergence behavior, but also end-task test performance.


\subsection{Effect of the Participation Ratio and Data Heterogeneity}
We next investigate the role of the participation ratio $r$ in order to illustrate the trade-off between update quality and communication efficiency, and to highlight the value of jointly optimizing device selection and power control. 

\begin{figure}[ht]
\centering
  \subfigure[Average training loss versus participation ratio $r$]{
    \includegraphics[width=3.2in]{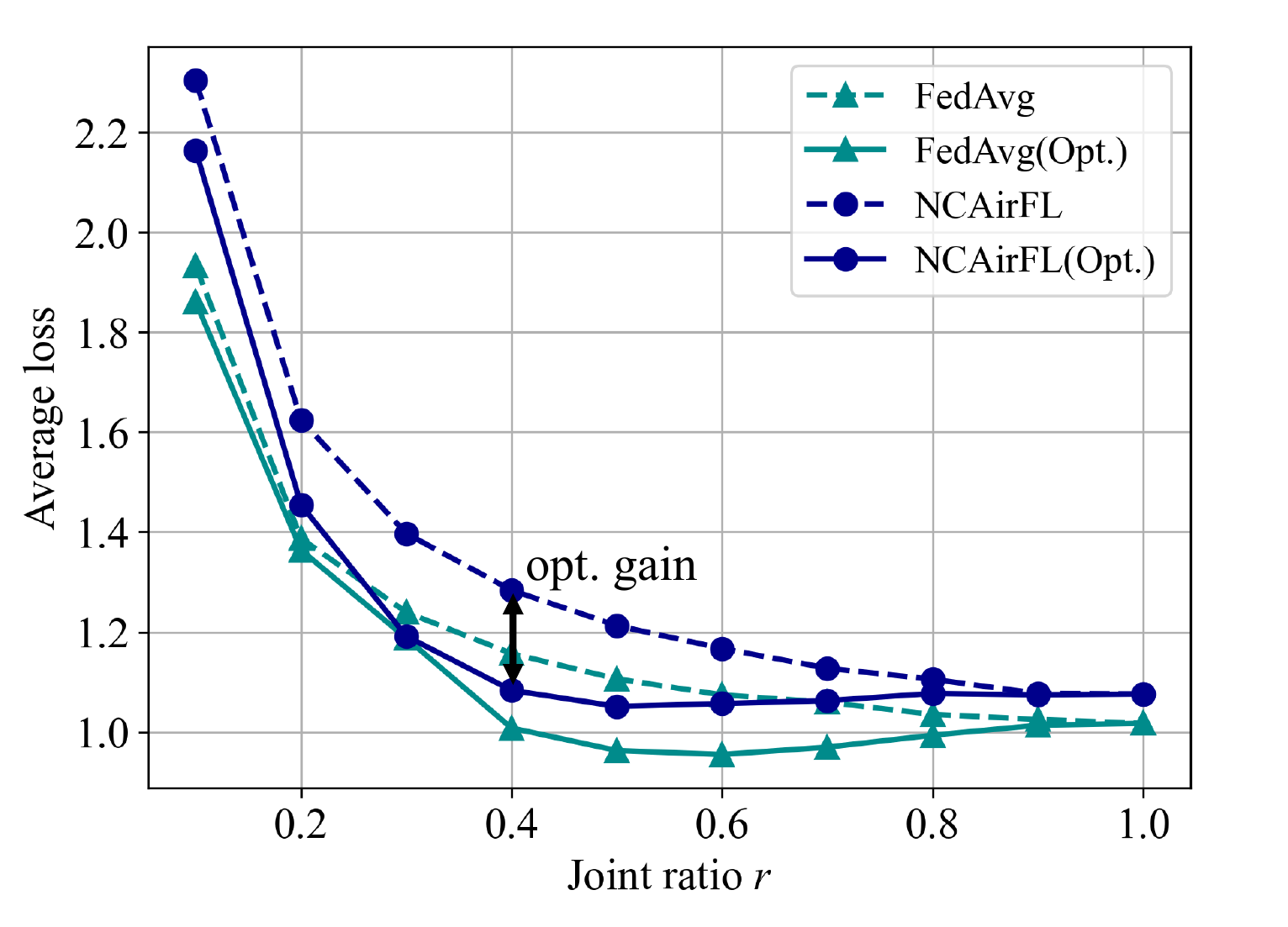}
    \label{fig:loss_vs_r_mnist}
  }
  \subfigure[Average test accuracy versus participation ratio $r$]{
    \includegraphics[width=3.2in]{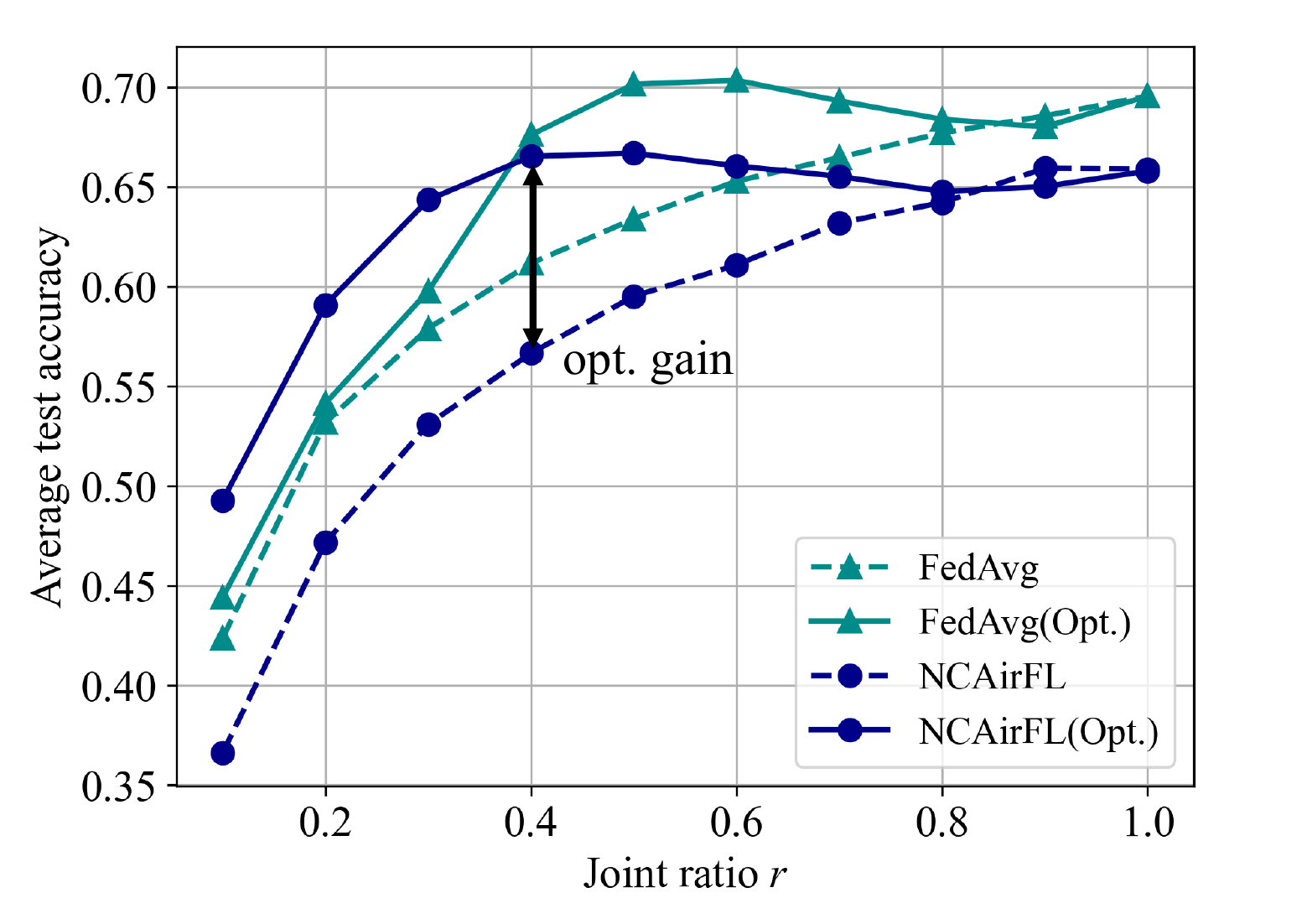}
    \label{fig:acc_vs_r_mnist}
  }
  \caption{Average training loss and test accuracy versus participation ratio $r$ for NCAirFL and FedAvg on MNIST. The average is taken over the communication rounds.}
\label{fig:jr result on MNIST}
\end{figure}

\begin{figure}[ht]
\centering

  \subfigure[Average training loss versus participation ratio $r$]{
    \includegraphics[width=3.2in]{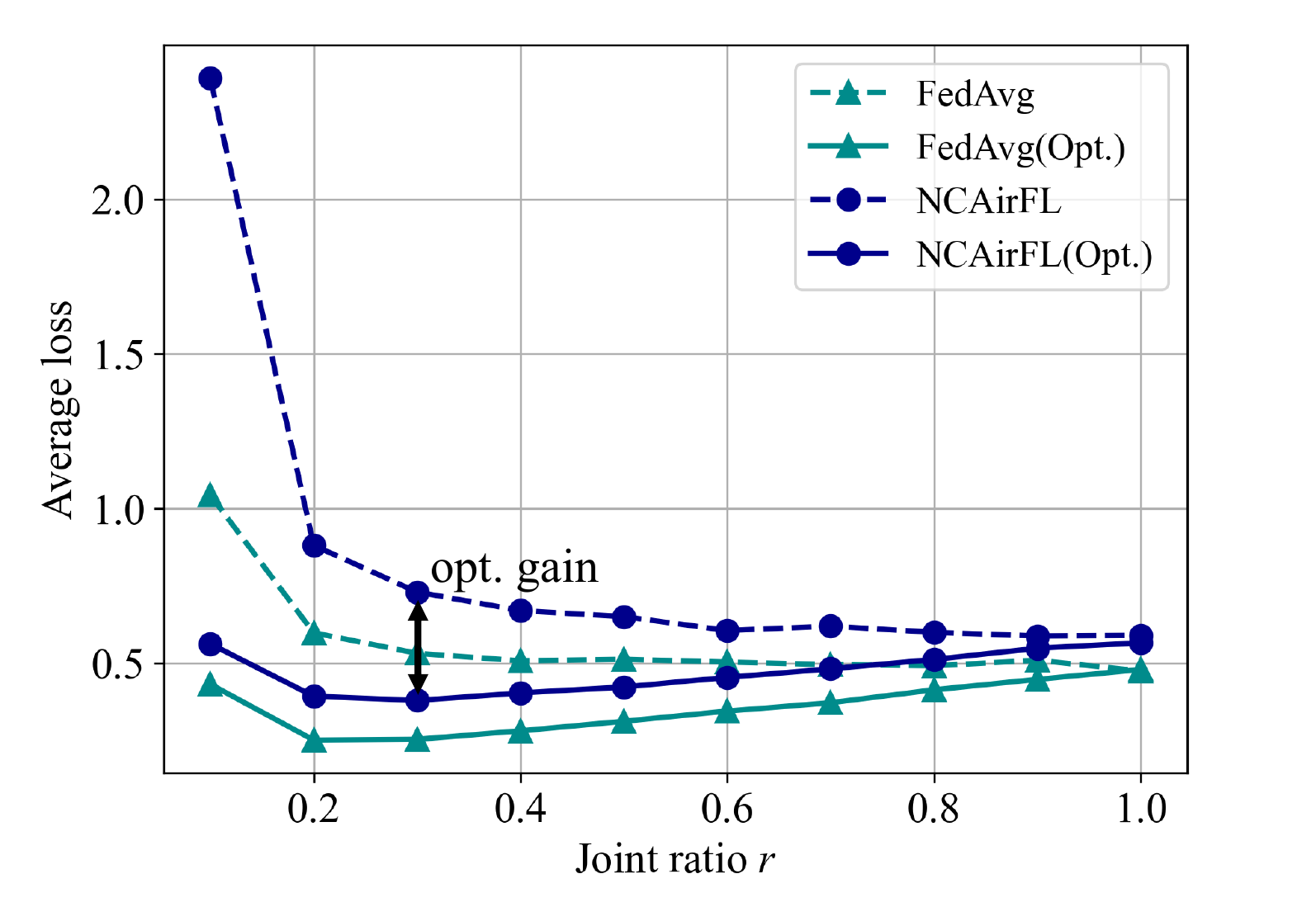}
    \label{fig:loss_vs_r_cifar10}
  }
  \subfigure[Average test accuracy versus participation ratio $r$]{
    \includegraphics[width=3.2in]{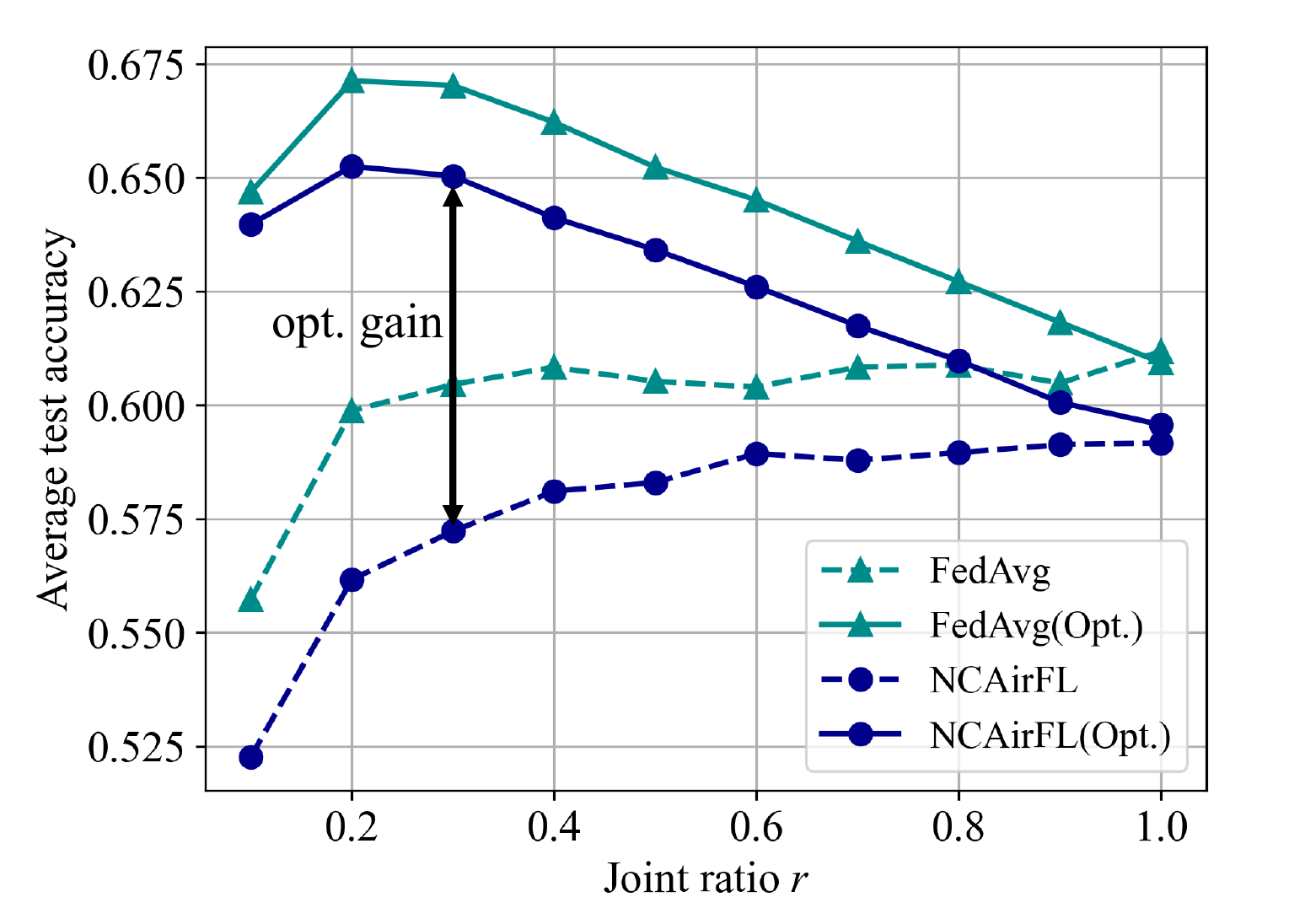}
    \label{fig:acc_vs_r_cifar10}
  }
  \caption{Average training loss and test accuracy versus participation ratio $r$ for NCAirFL and FedAvg on CIFAR-10. The average is taken over the communication rounds.}
  \label{fig:jr result on cifar10}
\end{figure}

Figs.~\ref{fig:loss_vs_r_mnist} and~\ref{fig:loss_vs_r_cifar10} show the communication-round-averaged training loss as functions of participation ratio $r$ for NCAirFL and FedAvg on MNIST and CIFAR-10, respectively.
For the non-optimized schemes, increasing $r$ generally improves the average training loss, since more devices contribute to each round. However, once optimized selection is introduced, the best performance is attained at an intermediate value of $r$ rather than at $r=1$. This phenomenon is visible for both datasets: on MNIST, the optimized schemes perform best around a moderate participation ratio, while on CIFAR-10 the gain is even more pronounced at relatively small values of $r$. The reason is that a smaller participation ratio gives the scheduler room to discard less useful or more power-limited devices, thereby improving the utility-noise trade-off captured by problem~\eqref{eq:selection optimization}. By contrast, when $r=1$, all devices must participate and the optimized and non-optimized schemes necessarily coincide.
The same conclusion is supported by the accuracy plots in Figs.~\ref{fig:acc_vs_r_mnist} and~\ref{fig:acc_vs_r_cifar10}: optimized selection improves the average accuracy over a broad range of participation ratios, with the largest gains obtained away from full participation.

% \subsection{Heterogeneity}
Finally, we study the effect of data heterogeneity on the benefit of optimized device selection by fixing the participation ratio to $r=0.5$ and varying the heterogeneity level $1/\varrho$ on MNIST.

\begin{figure}[ht]
\centering
  \subfigure[Average training loss versus the level of heterogeneity $1/\varrho$]{
    \includegraphics[width=3.2in]{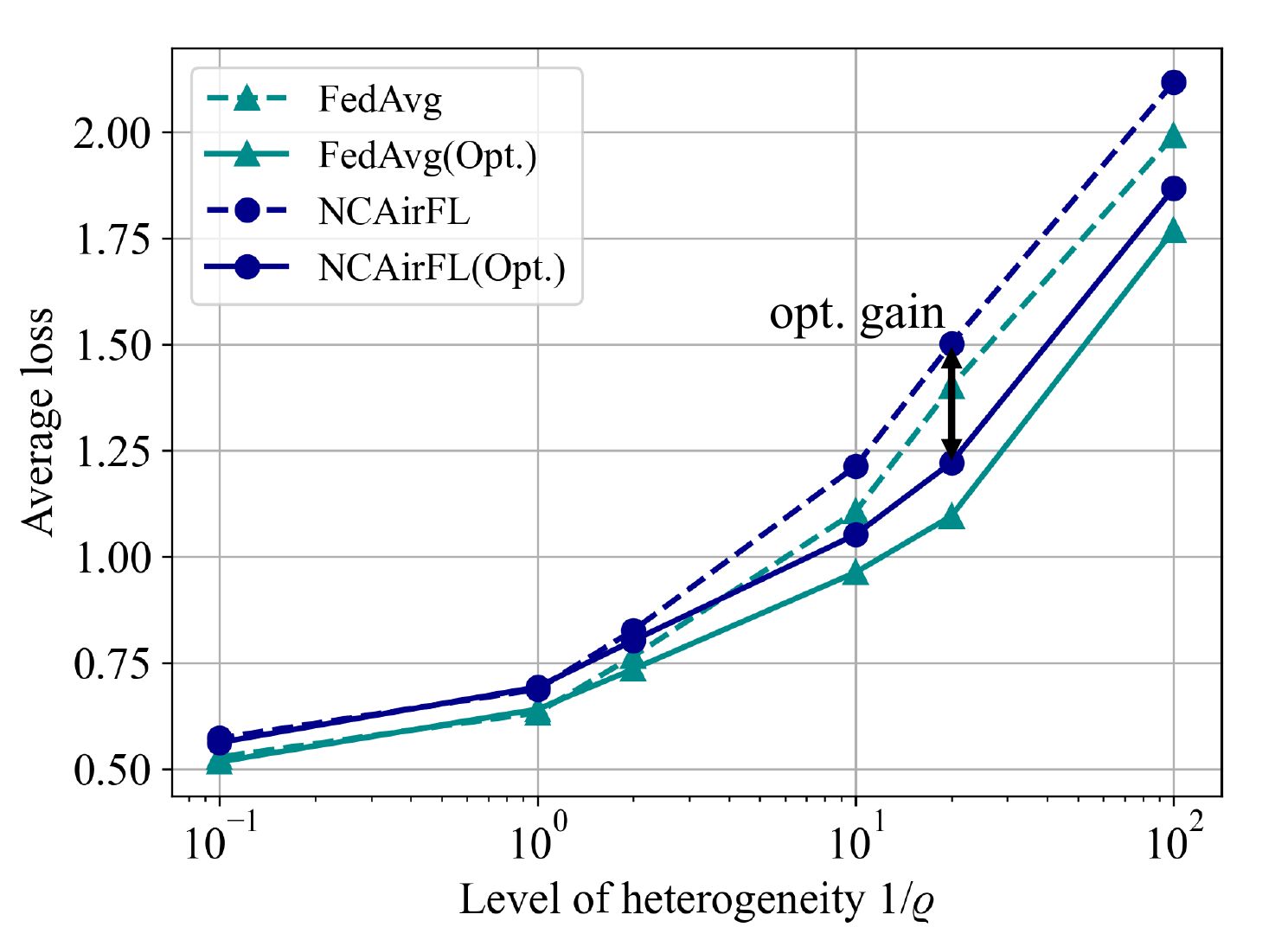}
    \label{fig:loss_vs_alpha_mnist}
  }
  \subfigure[Average test accuracy versus the level of heterogeneity $1/\varrho$]{
    \includegraphics[width=3.2in]{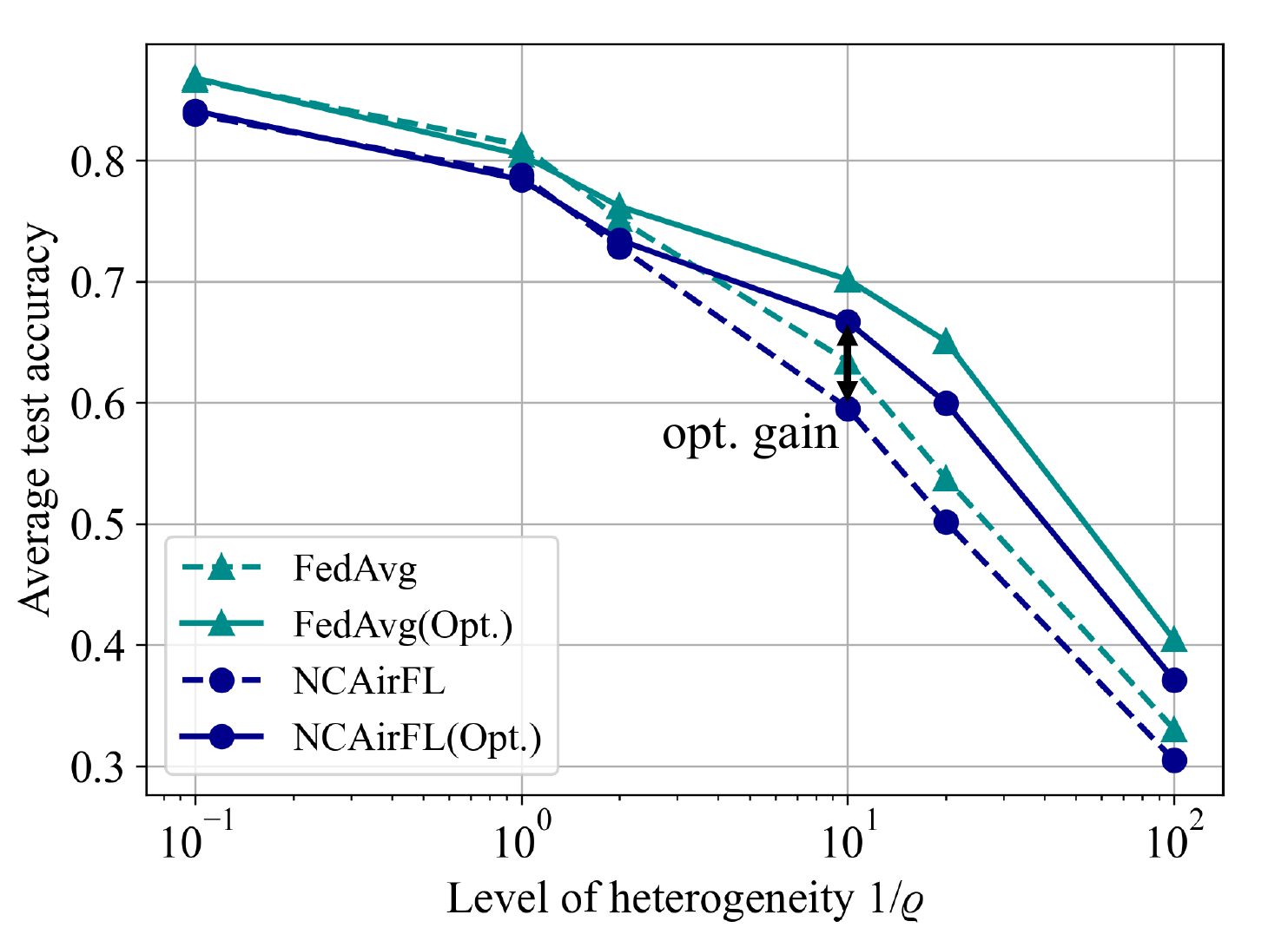}
    \label{fig:acc_vs_alpha_mnist}
  }
  \caption{Average training loss and test accuracy versus the level of heterogeneity $1/\varrho$ for NCAirFL and FedAvg on MNIST. The average is taken over the communication rounds.}
\label{fig:alpha result on MNIST}
\end{figure}

Figs.~\ref{fig:loss_vs_alpha_mnist} and~\ref{fig:acc_vs_alpha_mnist} show the corresponding average training loss and test accuracy.
As expected, increasing data heterogeneity makes the learning problem more difficult for all methods: the average loss increases, while the average accuracy decreases. At the same time, the benefit of optimized selection becomes more pronounced as heterogeneity grows. In particular, the gap between NCAirFL(Opt.) and NCAirFL is modest under weak heterogeneity, but it widens substantially in the highly heterogeneous regime. For example, at $1/\varrho=10$, the optimized variant improves the average test accuracy by about $10$ percentage points. This trend suggests that the proposed selection rule is especially valuable when client updates differ significantly in quality and statistical relevance.

\section{Conclusion}
This paper studied CSI semi-free AirFL for broadband single-antenna systems and proposed NCAirFL, which combines binary dithering, non-coherent aggregation, and memory-based error feedback to enable analog model aggregation without small-scale CSI at either the transmitters or the server. We established that NCAirFL retains the $\mathcal{O}(1/\sqrt{T})$ convergence rate for smooth non-convex objectives. To further improve communication efficiency, we derived a lower bound on the single-round reduction of the global loss and used it to formulate joint device selection and power control. The resulting surrogate problem admits a low-complexity optimal solution. Numerical results showed that NCAirFL approximates ideal FedAvg and that optimized device scheduling substantially accelerates convergence.

\appendix

\subsection{Proof Sketch of Lemma~\ref{lemma:contraction}}\label{appendix:sketch proof of contraction}
Define $\mv z_i^{(t)}=\mv m_i^{(t)}+\mv\Delta_i^{(t)}$ and condition on $(\mathcal F_t,\mv\Delta_i^{(t)})$, under which $\mv z_i^{(t)}$ is fixed. Expanding the square and using \eqref{eq:sparsified model different} gives
\begin{align}
&\mathbb{E}\left[\|\mv z_i^{(t)}-\mv\phi^{(t)}\odot\mv g_i^{(t)}\|^2\mid\mathcal F_t,\mv\Delta_i^{(t)}\right] \nonumber\\
% &=\|\mv z_i^{(t)}\|^2
% -2\mathbb{E}\left[\langle\mv\phi^{(t)}\odot\mv z_i^{(t)},\mv g_i^{(t)}\rangle\mid\mathcal F_t,\mv\Delta_i^{(t)}\right] \nonumber\\ 
% & \quad +\mathbb{E}\left[\|\mv\phi^{(t)}\odot\mv g_i^{(t)}\|^2\mid\mathcal F_t,\mv\Delta_i^{(t)}\right] \nonumber\\
&=\|\mv z_i^{(t)}\|^2-\mathbb{E}\left[\|\mv g_i^{(t)}\|^2\mid\mathcal F_t,\mv\Delta_i^{(t)}\right],
\label{eq:sketch contraction expansion}
\end{align}
where the last equality follows since the non-zero entries of $\mv g_i^{(t)}$ coincide with the positive entries of $\mv\phi^{(t)}\odot\mv z_i^{(t)}$ and $\|\mv\phi^{(t)}\odot\mv g_i^{(t)}\|^2=\|\mv g_i^{(t)}\|^2$. For coordinate $j$, define
\(
A_j=\Pr\left(z_{i,j}^{(t)}\phi_j^{(t)}>0\mid\mathcal F_t,\mv\Delta_i^{(t)}\right).
\)
If $z_{i,j}^{(t)}>0$, then $A_j=p$; if $z_{i,j}^{(t)}<0$, then $A_j=1-p$; and the contribution is zero if $z_{i,j}^{(t)}=0$. Hence, $A_j\ge\lambda=\min(p,1-p)$ whenever the corresponding squared entry is non-zero, and
\(
\mathbb{E}\!\left[\|\mv g_i^{(t)}\|^2\mid\mathcal F_t,\mv\Delta_i^{(t)}\right]
=\sum_{j=1}^{d}(z_{i,j}^{(t)})^2A_j
\ge\lambda\|\mv z_i^{(t)}\|^2.
\)
Substituting it into \eqref{eq:sketch contraction expansion} yields \eqref{eq:contraction}. Furthermore, $\lambda$ is maximized by $p=1/2$, which minimizes the contraction factor.

\subsection{Proof Sketch of Lemma~\ref{lemma:MSE of NC detection}}\label{appendix:sketch proof of MSE}
Recall that $e_j^{(t)}=r_j^{(t)}-(1/\rev{\eta})\sum_{i\in\mathcal I^{(t)}}g_{i,j}^{(t)}$, and define $\mathcal H_t=\sigma(\mathcal F_t,\mathcal I^{(t)},\{\mv\Delta_i^{(t)}\}_{i=1}^n,\mv\phi^{(t)})$. Conditional on $\mathcal H_t$, the transmitted quantities are fixed. Define
\(
    s_j^{(t)}=\sum_{i\in\mathcal I^{(t)}}h_{i,j}^{(t)}
    \sqrt{\Myfrac{g_{i,j}^{(t)}}{\rev{\eta}}},
\)
such that $y_j^{(t)}=\sqrt{\rho^{(t)}}s_j^{(t)}+n_j^{(t)}$. Expanding the square-law statistic yields
\begin{equation}
    e_j^{(t)}=a_j+b_j+c_j+d_j,
\end{equation}
where \(a_j=\sum_{i\in\mathcal I^{(t)}}(|h_{i,j}^{(t)}|^2-1)
    (\Myfrac{g_{i,j}^{(t)}}{\rev{\eta}}),\)
    \(b_j=(\Myfrac{2}{\rev{\eta}})\Re\{
    \sum_{i<k, (i,k\in\mathcal I^{(t)})}
    h_{i,j}^{(t)}h_{k,j}^{(t)*}
    \sqrt{g_{i,j}^{(t)}g_{k,j}^{(t)}}\},\)
    \(c_j=(\Myfrac{2}{\sqrt{\rho^{(t)}}})
    \Re\{n_j^{(t)*}s_j^{(t)}\},
    \)
    and 
    \(
    d_j=\Myfrac{(|n_j^{(t)}|^2-\sigma^2)}{\rho^{(t)}}
    \)
% \begin{align}
%     a_j&=\sum_{i\in\mathcal I^{(t)}}(|h_{i,j}^{(t)}|^2-1)
%     \frac{g_{i,j}^{(t)}}{\rev{\eta}}, \nonumber\\
%     b_j&=\frac{2}{\rev{\eta}}\Re\left\{
%     \sum_{\substack{i<k\\i,k\in\mathcal I^{(t)}}}
%     h_{i,j}^{(t)}h_{k,j}^{(t)*}
%     \sqrt{g_{i,j}^{(t)}g_{k,j}^{(t)}}\right\}, \nonumber\\
%     c_j&=\frac{2}{\sqrt{\rho^{(t)}}}
%     \Re\left\{n_j^{(t)*}s_j^{(t)}\right\},\qquad
%     d_j=\frac{|n_j^{(t)}|^2-\sigma^2}{\rho^{(t)}}.
%     \label{eq:sketch error components}
% \end{align}
Conditional on $\mathcal H_t$, the four terms \(a_j,b_j,c_j,\) and \(d_j\) have zero mean. By the independence and properness of the fading coefficients and the independence between the fading and noise, their mixed conditional second moments vanish. Therefore,
\(
    \mathbb{E}[(e_j^{(t)})^2\mid\mathcal H_t]
    =\mathbb{E}[a_j^2\mid\mathcal H_t]+\mathbb{E}[b_j^2\mid\mathcal H_t]
    \quad+\mathbb{E}[c_j^2\mid\mathcal H_t]+\mathbb{E}[d_j^2\mid\mathcal H_t].
\)
Using the channel moments and $n_j^{(t)}\sim\mathcal{CN}(0,\sigma^2)$, summing over $j\in[d]$, and applying Cauchy--Schwarz give
\begin{align}
    &\mathbb{E}[\|\mv e^{(t)}\|^2\mid\mathcal H_t]
    \le \frac{M_h}{\rev{\eta^2}}
    \sum_{i\in\mathcal I^{(t)}}\|\mv g_i^{(t)}\|^2
    +\frac{d\sigma^4}{(\rho^{(t)})^2}   \label{eq:sketch intermediate MSE} \\
    &+\frac{2\sqrt d\,\sigma^2}{\rev{\eta}\rho^{(t)}}
    \sum_{i\in\mathcal I^{(t)}}\|\mv g_i^{(t)}\| +\frac{1}{\rev{\eta^2}}
    \sum_{\substack{i\ne k\\i,k\in\mathcal I^{(t)}}}
    \left(\|\mv g_i^{(t)}\|^2
    +\|\mv g_k^{(t)}\|^2\right). \nonumber
\end{align}
Assumption~\ref{assumption:bounded variance} and \rev{the tower rule conditional on $\mathcal F_0$ give $\mathbb E[\|\mv\Delta_i^{(t)}\|^2\mid\mathcal F_0]\le\eta^2Q^2G^2$}. Together with the expected memory bound and \eqref{eq:sparsified model different}, this yields
\begin{align}
    \mathbb E[\|\mv g_i^{(t)}\|^2\mid\rev{\mathcal F_0}]
    &\le2\mathbb E[\|\mv m_i^{(t)}\|^2\mid\rev{\mathcal F_0}]+2\mathbb E[\|\mv\Delta_i^{(t)}\|^2\mid\rev{\mathcal F_0}]\nonumber\\
    &\le\eta^2\tilde G^2.
    \label{eq:sketch g bound}
\end{align}
Using $g_{i,j}^{(t)}\ge0$, the average-power constraint~\eqref{eq:power constraint}, and \eqref{eq:sketch g bound} gives
\begin{equation}
    \frac{1}{\rho^{(t)}}
    \le\frac{\tilde G}{\sqrt d\min_{i\in[n]}\kappa_iP_i}.
    \label{eq:sketch rho bound}
\end{equation}
Finally, \rev{applying the tower property to} \eqref{eq:sketch intermediate MSE} \rev{conditional on $\mathcal F_0$}, using \eqref{eq:sketch g bound}--\eqref{eq:sketch rho bound}, and defining $\rho_{\min}=\min_i\kappa_iP_i/\sigma^2$ give \eqref{eq:MSE bound}. Its four RHS terms correspond to the fading, noise--noise, inter-device, and signal--noise contributions, respectively.

\subsection{Proof Sketch of Proposition~\ref{proposition:convergence}}\label{appendix:sketch proof of convergence}
We use a perturbed-iterate argument. Define the virtual sequence $\{\tilde{\mv\theta}^{(t)}\}$ as
\begin{equation}
    {\tilde{\mv\theta}^{(t+1)}
    =\tilde{\mv\theta}^{(t)}
    +\frac{1}{s}\sum_{i\in\mathcal I^{(t)}}\mv\Delta_i^{(t)}
    +\frac{\eta}{s}\mv\phi^{(t)}\odot\mv e^{(t)},}
    \label{eq:sketch virtual sequence}
\end{equation}
with $\tilde{\mv\theta}^{(0)}=\mv\theta^{(0)}$. By \eqref{eq:conditional zero mean error} and the tower property, $\mathbb E[\mv\phi^{(t)}\odot\mv e^{(t)}\mid\mathcal F_t]=\mv 0$. From the global and memory updates, the real and virtual iterates satisfy
\begin{equation}
    {\mv\theta^{(t)}-\tilde{\mv\theta}^{(t)}
    =-\frac{1}{s}\sum_{i=1}^{n}\mv m_i^{(t)},}
    \label{eq:sketch real virtual relation}
    \end{equation}
    \rev{which follows by induction. 
    % It holds at $t=0$ because the real and virtual iterates coincide and all memories are zero. If it holds at round $t$, substituting the two iterate updates and applying \eqref{eq:memory update rule} gives the same identity at $t+1$.
    }
    \rev{Applying $L$-smoothness to \eqref{eq:sketch virtual sequence} and taking total expectation immediately gives}
    \begin{align}
    &\mathbb{E}\!\left[f(\tilde{\mv\theta}^{(t+1)})\right]
    \le \mathbb E\!\left[f(\tilde{\mv\theta}^{(t)})\right] +\mathbb{E}\!\left[\left\langle\nabla f(\tilde{\mv\theta}^{(t)}),
    \frac{1}{s}\sum_{i\in\mathcal I^{(t)}}\mv\Delta_i^{(t)}
    \right\rangle\right] \nonumber\\
    &\quad+\frac{L}{2}\mathbb{E}\!\left[\left\|
    \frac{1}{s}\sum_{i\in\mathcal I^{(t)}}\mv\Delta_i^{(t)}
    +\frac{\eta}{s}\mv\phi^{(t)}\odot\mv e^{(t)}
    \right\|^2\right].
    \label{eq:sketch smoothness convergence}
\end{align}
\rev{Because \(\mathcal{I}^{(t)}\) is sampled independently of the preceding initialized sets, it is independent of the iterate entering round $t$.} Uniform selection therefore makes the active-device average unbiased under total expectation. Under Assumptions~\ref{assumption: L-smoothness}--\ref{assumption:heterogeneity}, the local-SGD drift satisfies, for $\eta\le 1/(\sqrt{2}QL)$,
\begin{multline}
\mathbb E\!\left[\left\|\mv\theta_i^{(t,q)}-\mv\theta^{(t)}\right\|^2\middle|\mathcal F_t\right]
\le
5Q\eta^2\!\left(\sigma_l^2+6Q\sigma_g^2\right)\\
+30Q^2\eta^2\|\nabla f(\mv\theta^{(t)})\|^2.
\label{eq:sketch local update bound}
\end{multline}
Taking total expectation, using \eqref{eq:sketch real virtual relation} with \(\mathbb E[\|\mv m_i^{(t)}\|^2] \le (\Myfrac{4(1-\lambda^2)}{\lambda^2}) \eta^2 Q^2 G^2\), Jensen's inequality, Young's inequality, and $\eta\le1/(\sqrt{240}QL)$ gives
% \begin{align}
%     & \mathbb{E}\!\left[\left\langle\nabla f(\tilde{\mv\theta}^{(t)}),
%     \frac{1}{s}\sum_{i\in\mathcal I^{(t)}}\mv\Delta_i^{(t)}
%     \right\rangle\right] \nonumber\\
%     & \le-\frac{\eta Q}{4}\mathbb E\|\nabla f(\mv\theta^{(t)})\|^2
%     +5Q^2\eta^3L^2\!\left(\sigma_l^2+6Q\sigma_g^2\right) \nonumber\\
%     &\quad +30Q^3\eta^3L^2\mathbb E\|\nabla f(\mv\theta^{(t)})\|^2
%     +\frac{6Q^3L^2\eta^3(1-\lambda^2)}{r^2\lambda^2}G^2.
%     \label{eq:sketch first order convergence pre stepsize}
% \end{align}
% Since $\eta\le1/(\sqrt{240}QL)$, we have $30\eta^2Q^2L^2\le1/8$. Hence,
\begin{align}
    &\mathbb{E}\!\left[\left\langle\nabla f(\tilde{\mv\theta}^{(t)}),
    \frac{1}{s}\sum_{i\in\mathcal I^{(t)}}\mv\Delta_i^{(t)}
    \right\rangle\right] \nonumber\\
    &\le-\frac{\eta Q}{8}\mathbb E\|\nabla f(\mv\theta^{(t)})\|^2
    +5Q^2\eta^3L^2\!\left(\sigma_l^2+6Q\sigma_g^2\right) \nonumber\\
    &\quad +\frac{6Q^3L^2\eta^3(1-\lambda^2)}{r^2\lambda^2}G^2.
    \label{eq:sketch first order convergence}
\end{align}
\rev{For the second-order term in \eqref{eq:sketch smoothness convergence}, condition first on $(\mathcal F_t,\mathcal I^{(t)},\{\mv\Delta_i^{(t)}\}_{i=1}^n,\mv\phi^{(t)})$. The innovation average and $\mv\phi^{(t)}$ are then fixed, while \eqref{eq:conditional zero mean error} makes the cross term zero. Moreover, $\|s^{-1}\sum_i\mv\Delta_i^{(t)}\|^2\le s^{-1}\sum_i\|\mv\Delta_i^{(t)}\|^2$, $\mathbb E\|\mv\Delta_i^{(t)}\|^2\le\eta^2Q^2G^2$, and $\|\mv\phi^{(t)}\odot\mv e^{(t)}\|=\|\mv e^{(t)}\|$. Hence,}
\begin{equation}
\frac{L}{2}\mathbb{E}\left\|
\frac{1}{s}\sum_{i\in\mathcal I^{(t)}}\mv\Delta_i^{(t)}
+\frac{\eta}{s}\mv\phi^{(t)}\odot\mv e^{(t)}
\right\|^2
\le\frac{L\eta^2Q^2G^2}{2}
+\frac{\eta^2LG_e^2}{2s^2},
\label{eq:sketch second order convergence}
\end{equation}
where Lemma~\ref{lemma:MSE of NC detection} and the tower property are used in the last term. Substituting \eqref{eq:sketch first order convergence} and \eqref{eq:sketch second order convergence} into \eqref{eq:sketch smoothness convergence}, summing over $t=0,\ldots,T-1$, and using $f(\tilde{\mv\theta}^{(T)})\ge f_*$ yield \eqref{eq:convergence bound}.

\subsection{Proof Sketch of Proposition~\ref{proposition:single-round reduction}}\label{appendix:sketch proof of single-round reduction}
\rev{Let $\mathcal I^{(t)}$ and $\rho^{(t)}>0$ be $\mathcal F_0$-measurable. Throughout this proof sketch, every displayed expectation is conditional on $\mathcal F_0$. Hence, $\mathcal I^{(t)}$ and $\rho^{(t)}$ are fixed inside these expectations.} Applying $L$-smoothness to the NCAirFL update gives
\begin{align}
& {\mathbb{E}\!\left[f(\mv\theta^{(t+1)})\right]}
\le {\mathbb E\!\left[f(\mv\theta^{(t)})\right]} +\mathbb{E}\!\left[\left\langle\nabla f(\mv\theta^{(t)}),
\frac{1}{s}\sum_{i\in\mathcal I^{(t)}}
\mv\phi^{(t)}\odot\mv g_i^{(t)}
\right\rangle\right] \nonumber\\
&+{\frac{L}{2}\mathbb{E}\!\left[\left\|
\frac{1}{s}\sum_{i\in\mathcal I^{(t)}}
\mv\phi^{(t)}\odot\mv g_i^{(t)}
+\frac{\eta}{s}\mv\phi^{(t)}\odot\mv e^{(t)}
\right\|^2\right].}
\label{eq:sketch smoothness single round}
\end{align}
To bound the first-order term, for each $i\in\mathcal I^{(t)}$, decompose
\(
-\eta Q\nabla f(\mv\theta^{(t)})
=-\mv A_i-\mv B_i-\mv C_i+\mv D_i
+\mv\phi^{(t)}\odot\mv g_i^{(t)},
\)
{where $\mv A_i$ is the local-model drift term, $\mv B_i$ is the data-heterogeneity term, $\mv C_i$ is the stochastic-gradient error, and $\mv D_i=\mv m_i^{(t+1)}-\mv m_i^{(t)}$ is the memory increment. Since $\eta\le1/(\sqrt{2}QL)$, the local-update bound in \eqref{eq:sketch local update bound} applies. Together with Assumptions~\ref{assumption:bounded variance}--\ref{assumption:heterogeneity}, $\mathbb E\|\nabla f(\mv\theta^{(t)})\|^2\le G^2$, and the expected memory bound, it gives}
\(
\mathbb{E}\|\mv A_i\|^2
\le
5\eta^4L^2Q^3\!\left(\sigma_l^2+6Q\sigma_g^2\right)
+30\eta^4L^2Q^4G^2,
\)
\(
\mathbb{E}\|\mv B_i\|^2 \le\eta^2Q^2\sigma_g^2,
\)
\(
\mathbb{E}\|\mv C_i\|^2\le\eta^2Q\sigma_l^2,
\)
and 
\( 
\mathbb{E}\|\mv D_i\|^2 \le (\Myfrac{(16\eta^2(1-\lambda^2))}{\lambda^2})Q^2G^2.
\)
% \begin{align}
% {\mathbb{E}\|\mv A_i\|^2}
% &{\le
% 5\eta^4L^2Q^3\!\left(\sigma_l^2+6Q\sigma_g^2\right)
% +30\eta^4L^2Q^4G^2,} \nonumber\\
% {\mathbb{E}\|\mv B_i\|^2}&{\le\eta^2Q^2\sigma_g^2,\qquad
% \mathbb{E}\|\mv C_i\|^2\le\eta^2Q\sigma_l^2,} \nonumber\\
% {\mathbb{E}\|\mv D_i\|^2}
% &{\le\frac{16\eta^2(1-\lambda^2)}{\lambda^2}Q^2G^2.}
% \label{eq:sketch decomposition bounds}
% \end{align}
Using the above bounds, Cauchy--Schwarz, and $\|\mv\phi^{(t)}\odot\mv x\|=\|\mv x\|$, we obtain
\begin{align}
&{\mathbb{E}\!\left[\left\langle\nabla f(\mv\theta^{(t)}),
\frac{1}{s}\sum_{i\in\mathcal I^{(t)}}
\mv\phi^{(t)}\odot\mv g_i^{(t)}
\right\rangle\right]} \nonumber\\
&{\le-\frac{1}{s\eta Q}\sum_{i\in\mathcal I^{(t)}}
\mathbb{E}\|\mv g_i^{(t)}\|^2
+\frac{C_1}{s}\sum_{i\in\mathcal I^{(t)}}
\sqrt{\mathbb{E}\|\mv g_i^{(t)}\|^2}.}
\label{eq:sketch first order single round}
\end{align}
{For the second-order term, the detector's conditional zero-mean property eliminates the cross term before total expectation is taken. The intermediate MSE bound obtained in the proof of Lemma~\ref{lemma:MSE of NC detection} then yields}
\begin{align}
&\frac{L}{2}\mathbb{E}\left\|
\frac{1}{s}\sum_{i\in\mathcal I^{(t)}}
\mv\phi^{(t)}\odot\mv g_i^{(t)}
+\frac{\eta}{s}\mv\phi^{(t)}\odot\mv e^{(t)}
\right\|^2 \nonumber\\
&\le\frac{L}{2s}\sum_{i\in\mathcal I^{(t)}}
\mathbb{E}\|\mv g_i^{(t)}\|^2
+\frac{\eta^2L}{2s^2}
\left((M_h+2s-2)s\tilde G^2\right. \nonumber\\
&\hspace{1.0in}\left.
+\frac{d\sigma^4}{(\rho^{(t)})^2}
    +\frac{2s\sqrt d\,\sigma^2}{\rho^{(t)}}\tilde G\right).
\label{eq:sketch second order single round}
\end{align}
Finally, substituting \eqref{eq:sketch first order single round} and \eqref{eq:sketch second order single round} into \eqref{eq:sketch smoothness single round} and rearranging gives
\eqref{eq:single-round reduction}.

\bibliographystyle{IEEEtran}
\bibliography{ref}

@ARTICLE{11475389,
  author={Ul Abrar, Muhammad Faraz and Michelusi, Nicolò},
  journal={IEEE Trans. Wireless Commun.}, 
  title={Biased Federated Learning Under Wireless Heterogeneity}, 
  year={2026},
  volume={25},
  number={},
  pages={16449-16462},
  doi={10.1109/TWC.2026.3678173}}

@INPROCEEDINGS{11587673,
  author={Michelusi, Nicolò},
  booktitle={ICC 2026 - IEEE International Conference on Communications}, 
  title={Interference-Robust Non-Coherent Over-the-Air Computation for Decentralized Optimization}, 
  year={2026},
  address={Glasgow, United Kingdom},
  month={May}
  }

@STRING{ieee="IEEE Computer Society Press"}

@STRING{IEEE_J_IT         = "{IEEE} Trans. Inf. Theory"}

@STRING{IEEE_J_JSAC       = "{IEEE} J. Sel. Areas Commun."}

@STRING{IEEE_J_SP         = "{IEEE} Trans. Signal Process."}

@STRING{IEEE_J_WCOM       = "{IEEE} Trans. Wireless Commun."}

@inproceedings{mcmahan2017communication,
  title={Communication-efficient learning of deep networks from decentralized data},
  author={McMahan, Brendan and Moore, Eider and Ramage, Daniel and Hampson, Seth and y Arcas, Blaise Aguera},
  booktitle={Proc. Artificial Intelligence and Statistics},
  pages={},
  address = {FL, USA},
  year={2017},
  month = {Apr.}
}

@article{Tao2024Federated,
  author = {M. Tao and others},
  title = {Federated Edge Learning for {6G}: Foundations, Methodologies, and Applications},
  journal = {Proc. IEEE},
  pages = {1--39},
  year = {2024},
}

@article{cao2021optimized,
  title={Optimized power control design for over-the-air federated edge learning},
  author={Cao, Xiaowen and Zhu, Guangxu and Xu, Jie and Wang, Zhiqin and Cui, Shuguang},
  journal=IEEE_J_JSAC,
  volume={40},
  number={1},
  pages={342--358},
  year={2021},
  publisher={IEEE}
}

@inproceedings{yang2021achieving,
  title={Achieving linear speedup with partial worker participation in non-iid federated learning},
  author={Yang, Haibo and Fang, Minghong and Liu, Jia},
  booktitle={Proc. International Conference on Learning Representations (ICLR)},
  year={2021}
}

@article{yang2020federated,
  title={Federated learning via over-the-air computation},
  author={Yang, Kai and Jiang, Tao and Shi, Yuanming and Ding, Zhi},
  journal=IEEE_J_WCOM,
  volume={19},
  number={3},
  pages={2022--2035},
  year={2020},
  publisher={IEEE}
}

@article{amiri2020federated,
  title={Federated learning over wireless fading channels},
  author={Amiri, Mohammad Mohammadi and G{\"u}nd{\"u}z, Deniz},
  journal=IEEE_J_WCOM,
  volume={19},
  number={5},
  pages={3546--3557},
  year={2020},
  publisher={IEEE}
}

@article{nazer2007computation,
  title={Computation over multiple-access channels},
  author={Nazer, Bobak and Gastpar, Michael},
  journal=IEEE_J_IT,
  volume={53},
  number={10},
  pages={3498--3516},
  year={2007},
  publisher={IEEE}
}

@article{tegin2023federated,
  title={Federated Learning with Over-the-Air Aggregation over Time-Varying Channels},
  author={Tegin, Busra and Duman, Tolga M},
  journal=IEEE_J_WCOM,
  year={2023},
  publisher={IEEE}
}

@article{sery2021over,
  title={Over-the-air federated learning from heterogeneous data},
  author={Sery, Tomer and Shlezinger, Nir and Cohen, Kobi and Eldar, Yonina C},
  journal={IEEE Trans. Signal Process.},
  volume={69},
  pages={3796--3811},
  year={2021},
  publisher={IEEE}
}

@inproceedings{basu19qsparse,
	title={{Qsparse-local-SGD}: Distributed {SGD} with quantization, sparsification and local computations},
	author={Basu, Debraj and Data, Deepesh and Karakus, Can and Diggavi, Suhas},
	booktitle={Proc. Advances in Neural Information Processing Systems},
	address={Vancouver, Canada},
	month=dec,
	year={2019}
}

@article{sery2020analog,
  title={On analog gradient descent learning over multiple access fading channels},
  author={Sery, Tomer and Cohen, Kobi},
  journal=IEEE_J_SP,
  volume={68},
  pages={2897--2911},
  year={2020},
  publisher={IEEE}
}

@article{yang2021revisiting,
  title={Revisiting analog over-the-air machine learning: The blessing and curse of interference},
  author={Yang, Howard H and Chen, Zihan and Quek, Tony QS and Poor, H Vincent},
  journal={ IEEE J Sel. Top. Signal Process.},
  volume={16},
  number={3},
  pages={406--419},
  year={2021},
  publisher={IEEE}
}

@article{zhu2019broadband,
  author={Zhu, Guangxu and Wang, Yong and Huang, Kaibin},
  journal=IEEE_J_WCOM, 
  title={Broadband Analog Aggregation for Low-Latency Federated Edge Learning}, 
  year={2020},
  volume={19},
  number={1},
  pages={491-506},
  doi={10.1109/TWC.2019.2946245}}

@article{amiri2021blind,
  title={Blind federated edge learning},
  author={Amiri, Mohammad Mohammadi and Duman, Tolga M and G{\"u}nd{\"u}z, Deniz and Kulkarni, Sanjeev R and Poor, H Vincent},
  journal=IEEE_J_WCOM,
  volume={20},
  number={8},
  pages={5129--5143},
  year={2021},
  publisher={IEEE}
}

@article{wei2023random,
  title={Random orthogonalization for federated learning in massive {MIMO} systems},
  author={Wei, Xizixiang and Shen, Cong and Yang, Jing and Poor, H Vincent},
  journal=IEEE_J_WCOM,
  year={2023},
  publisher={IEEE}
}

@article{yue2022efficient,
  title={Efficient federated meta-learning over multi-access wireless networks},
  author={Yue, Sheng and Ren, Ju and Xin, Jiang and Zhang, Deyu and Zhang, Yaoxue and Zhuang, Weihua},
  journal={IEEE J. Sel. Areas Commun.},
  volume={40},
  number={5},
  pages={1556--1570},
  year={2022},
  publisher={IEEE}
}

@misc{lecun1998mnist,
  title={The {MNIST} database of handwritten digits},
  author={LeCun, Yann},
  url={http://yann.lecun.com/exdb/mnist/},
  year={1998}
}

@misc{krizhevsky2009cifar10,
  title={Learning multiple layers of features from tiny images},
  author={Krizhevsky, Alex and Hinton, Geoffrey and others},
  url={https://www.cs.toronto.edu/~kriz/learning-features-2009-TR.pdf},
  year={2009}
}

@inproceedings{yurochkin2019bayesian,
  title={Bayesian nonparametric federated learning of neural networks},
  author={Yurochkin, Mikhail and Agarwal, Mayank and Ghosh, Soumya and Greenewald, Kristjan and Hoang, Nghia and Khazaeni, Yasaman},
  booktitle={Proc. International conference on machine learning},
  address = {CA, USA},
  month = {Jun.},
  year={2019},
}

@article{kingma2014adam,
  title={Adam: A method for stochastic optimization},
  author={Kingma, Diederik P},
  journal={arXiv preprint arXiv:1412.6980},
  year={2014}
}

@inproceedings{he2016deep,
  title={Deep residual learning for image recognition},
  author={He, Kaiming and Zhang, Xiangyu and Ren, Shaoqing and Sun, Jian},
  booktitle={Proc. IEEE Conference on Computer Vision and Pattern Recognition},
  month={Jun.},
  address={Nevada, USA},
  year={2016}
}

@INPROCEEDINGS{wen2024AirFL-Mem,
  author={Wen, Haifeng and Xing, Hong and Simeone, Osvaldo},
  booktitle={2024 IEEE Wireless Communications and Networking Conference (WCNC)}, 
  title={{AirFL-Mem}: Improving Communication-Learning Trade-Off by Long-Term Memory}, 
  year={2024},
  month={Apr.},
  address={Dubai, United Arab Emirates},
}

@ARTICLE{michelusi2024non,
  author={Michelusi, Nicolò},
  journal=IEEE_J_SP, 
  title={Non-Coherent Over-the-Air Decentralized Gradient Descent}, 
  year={2024},
  volume={72},
  number={},
  pages={4618-4634},
  doi={10.1109/TSP.2024.3460690}}

@article{chen2018over,
  title={Over-the-air computation for {IoT} networks: Computing multiple functions with antenna arrays},
  author={Chen, Li and Zhao, Nan and Chen, Yunfei and Yu, F Richard and Wei, Guo},
  journal={IEEE Internet Things J.},
  volume={5},
  number={6},
  pages={5296--5306},
  year={2018},
  publisher={IEEE}
}

@article{dong2020blind,
  title={Blind over-the-air computation and data fusion via provable wirtinger flow},
  author={Dong, Jialin and Shi, Yuanming and Ding, Zhi},
  journal=IEEE_J_SP,
  volume={68},
  pages={1136--1151},
  year={2020},
  publisher={IEEE}
}

@ARTICLE{choi2022communication,
  author={Choi, Jinho},
  journal={IEEE J. Sel. Areas Inf. Theory},
  title={Communication-Efficient Distributed {SGD} Using Random Access for Over-the-Air Computation}, 
  year={2022},
  volume={3},
  number={2},
  pages={206-216},
}

@ARTICLE{ren2020scheduling,
  author={Ren, Jinke and He, Yinghui and Wen, Dingzhu and Yu, Guanding and Huang, Kaibin and Guo, Dongning},
  journal=IEEE_J_WCOM, 
  title={Scheduling for Cellular Federated Edge Learning With Importance and Channel Awareness}, 
  year={2020},
  volume={19},
  number={11},
  pages={7690-7703},
}

@ARTICLE{nguyen2021fast,
  author={Nguyen, Hung T. and Sehwag, Vikash and Hosseinalipour, Seyyedali and Brinton, Christopher G. and Chiang, Mung and Vincent Poor, H.},
  journal=IEEE_J_JSAC, 
  title={Fast-Convergent Federated Learning}, 
  year={2021},
  volume={39},
  number={1},
  pages={201-218}
}

@ARTICLE{sun2024channel,
  author={Sun, Yuchang and Lin, Zehong and Mao, Yuyi and Jin, Shi and Zhang, Jun},
  journal=IEEE_J_WCOM, 
  title={Channel and Gradient-Importance Aware Device Scheduling for Over-the-Air Federated Learning}, 
  year={2024},
  volume={23},
  number={7},
  pages={6905-6920} 
}

@ARTICLE{su2022data,
  author={Su, Liqun and Lau, Vincent K. N.},
  journal={IEEE Internet Things J.}, 
  title={Data and Channel-Adaptive Sensor Scheduling for Federated Edge Learning via Over-the-Air Gradient Aggregation}, 
  year={2022},
  volume={9},
  number={3},
  pages={1640-1654} 
}

@ARTICLE{du2023gradient,
  author={Du, Jun and Jiang, Bingqing and Jiang, Chunxiao and Shi, Yuanming and Han, Zhu},
  journal=IEEE_J_JSAC, 
  title={Gradient and Channel Aware Dynamic Scheduling for Over-the-Air Computation in Federated Edge Learning Systems}, 
  year={2023},
  volume={41},
  number={4},
  pages={1035-1050},
}

@misc{ITU-R2023Framework,
  author = {ITU-R},
  title = {Framework and Overall Objectives of the Future Development of {IMT} for 2030 and Beyond},
  year = {2023},
  url = {https://techblog.comsoc.org/2023/01/29/},
  organization = {ITU-R},
}

@article{ghadimi2013stochastic,
  title={Stochastic first-and zeroth-order methods for nonconvex stochastic programming},
  author={Ghadimi, Saeed and Lan, Guanghui},
  journal={SIAM journal on optimization},
  volume={23},
  number={4},
  pages={2341--2368},
  year={2013},
  publisher={SIAM}
}

@ARTICLE{wen2026tccn,
  author={Wen, Haifeng and Xing, Hong and Simeone, Osvaldo},
  journal={IEEE Trans. Cogn. Commun. Netw.}, 
  title={Pre-Training and Personalized Fine-Tuning via Over-the-Air Federated Meta-Learning: Convergence-Generalization Trade-Offs}, 
  year={2026},
  volume={12},
  number={},
  pages={4911-4925}
  }

@INPROCEEDINGS{deng2025robust,
  author={Deng, Yuhang and Chen, Zheng and Larsson, Erik G.},
  booktitle={ICC 2025 - IEEE International Conference on Communications}, 
  title={Robust and Efficient Average Consensus with Non-Coherent Over-the-Air Aggregation}, 
  year={2025},
  address={Montreal, QC, Canada},
  month={Jun.}
  }

@book{cover1991elements,
  title={Elements of information theory},
  author={Cover, Thomas M and Thomas, Joy A and Kieffer, John},
  volume={2},
  year={1991},
  publisher={wiley New York}
}

@techreport{3GPP,
  author       = {{3GPP}},
  title        = {{NR}; Physical channels and modulation},
  institution  = {3rd Generation Partnership Project (3GPP)},
  number       = {TS 38.211},
  version      = {V16.7.0},
  year         = {2021},
  month        = sep,
  note         = {Release 16},
  type         = {Technical Report}
}

@INPROCEEDINGS{wen2025icc,
  author={Wen, Haifeng and Michelusi, Nicolò and Simeone, Osvaldo and Xing, Hong},
  booktitle={ICC 2025 - IEEE International Conference on Communications}, 
  title={NCAirFL: CSI-Free Over-the-Air Federated Learning Based on Non-Coherent Detection}, 
  year={2025},
  month={Jun.},
  address={Montreal, QC, Canada}
  }

\end{document}